\documentclass[a4paper,fleqn]{cas-sc}

\usepackage{acronym} 

\acrodef{A.I.}[AI]{artificial intelligence}
\acrodef{SEM}[SEM]{scanning electron microscopy}
\acrodef{NSD}[NSD]{Normalized Surface Distance}
\acrodef{clDice}[clDice]{centerline Dice}
\acrodef{TV}[TV]{Total Variation}
\acrodef{CNN}[CNN]{convolutional neural network}
\acrodef{PDE}[PDE]{partial differential equation}
\acrodef{PDX}[PDX]{patient-derived xenograft}
\acrodef{SBFSEM}[SBF-SEM]{serial block face-SEM}
\acrodef{GAN}[GAN]{Generative Adversarial Network}
\acrodef{SPADE}[SPADE]{SPatially Adaptative (DE)normalization}
\acrodef{FT}[FT]{fine tuning}
\acrodef{COp-Net}[COp-Net]{Closing Operator}
\acrodef{FOV}[FOV]{field of view}

\usepackage{graphicx}
\usepackage{subcaption}
\usepackage{tabularx}
\usepackage{booktabs}
\usepackage{multirow}
\usepackage{array,makecell}

\newcolumntype{C}[1]{>{\centering\arraybackslash}m{#1}} %
\newcolumntype{L}[1]{>{\raggedright}m{#1}}

\usepackage[authoryear]{natbib}

\def\tsc#1{\csdef{#1}{\textsc{\lowercase{#1}}\xspace}}
\tsc{WGM}
\tsc{QE}
\tsc{EP}
\tsc{PMS}
\tsc{BEC}
\tsc{DE}

\begin{document}
\let\WriteBookmarks\relax
\def\floatpagepagefraction{1}
\def\textpagefraction{.001}
\shorttitle{Enhanced Cell Segmentation in SEM Images}
\shortauthors{F. Robert et~al.}

\title [mode = title]{Enhancing Cell Instance Segmentation in Scanning Electron Microscopy Images via a Deep Contour Closing Operator}                      
%
%

\author[label0,label1,label2]{Florian Robert}
\fnref{cor2}
\corref{cor1}
\ead{florian.robert@math.u-bordeaux.fr}
\fntext[cor2]{Corresponding author}
\cortext[cor1]{These two authors equally contributed to this work}

\affiliation[label0]{organization={Univ. of Bordeaux, CNRS, Institut de Mathématiques de Bordeaux, IMB, UMR5251},
           addressline={351 cours de la Libération}, 
           city={Talence},
           postcode={F-33400}, 
          country={France}}
          
\affiliation[label1]{organization={INRIA Bordeaux, MONC team},
           addressline={200 avenue de la Vieille Tour}, 
           city={Talence},
           postcode={F-33400 }, 
          country={France}}

\credit{Conceptualization, Methodology, Software, Validation, Writing - Original Draft}

\author[label2,label3]{Alexia Calovoulos}
\corref{cor1}
\ead{alexia.calovoulos@u-bordeaux.fr}

\credit{Resources, Data curation}

\affiliation[label2]{organization={Univ. Bordeaux, INSERM, Bordeaux Institute in Oncology, BRIC, U1312, MIRCADE team},
            addressline={146 rue Léo Saignat}, 
            city={Bordeaux},
            postcode={33000}, 
            country={France}}
            
\affiliation[label3]{organization={Univ. Bordeaux, CNRS, INSERM, Bordeaux Imaging Centre, BIC, UAR 3420, US 4},
            addressline={146 rue Léo Saignat}, 
            city={Bordeaux},
            postcode={33000}, 
            country={France}}
            
\author[label0]{Laurent Facq}
\ead{laurent.facq@math.u-bordeaux.fr} 
\credit{Software}

\author[label3]{Fanny Decoeur}
\ead{fanny.decoeur@u-bordeaux.fr}
\credit{Resources, Supervision}
            
\author[label3]{Etienne Gontier}
\ead{etienne.gontier@u-bordeaux.fr}
\credit{Supervision, Project administration, Funding acquisition}

\author[label2]{Christophe F. Grosset}
\ead{christophe.grosset@inserm.fr}
\credit{Supervision, Project administration, Funding acquisition, Writing - Review}
       
\author[label0,label1]{Baudouin {Denis de Senneville}}
\ead{bdenisde@math.u-bordeaux.fr}     
\credit{Supervision, Project administration, Funding acquisition, Writing - Review \& Editing}

\begin{abstract}
Accurately segmenting and individualizing cells in \ac{SEM} images is a highly promising technique for elucidating tissue architecture in oncology. While current \ac{A.I.}-based methods are effective, errors persist, necessitating time-consuming manual corrections, particularly in areas where the quality of cell contours in the image is poor and requires gap filling.

This study presents a novel \ac{A.I.}-driven approach for refining cell boundary delineation to improve instance-based cell segmentation in \ac{SEM} images, also reducing the necessity for residual manual correction. A \ac{CNN} \ac{COp-Net} is introduced to address gaps in cell contours, effectively filling in regions with deficient or absent information. 
The network takes as input cell contour probability maps with potentially inadequate or missing information and outputs corrected cell contour delineations. The lack of training data was addressed by generating low integrity probability maps using a tailored \ac{PDE}. To ensure reproducibility, \ac{COp-Net} weights and the source code for solving the \ac{PDE} are publicly available at \url{https://github.com/Florian-40/CellSegm}.

We showcase the efficacy of our approach in augmenting cell boundary precision using both private \ac{SEM} images from \ac{PDX} hepatoblastoma tissues and publicly accessible images datasets. The proposed cell contour closing operator exhibits a notable improvement in tested datasets, achieving respectively close to $50\%$ (private data) and $10\%$ (public data) increase in the accurately-delineated cell proportion compared to state-of-the-art methods. Additionally, the need for manual corrections was significantly reduced, therefore facilitating the overall digitalization process.

Our results demonstrate a notable enhancement in the accuracy of cell instance segmentation, particularly in highly challenging regions where image quality compromises the integrity of cell boundaries, necessitating gap filling. Therefore, our work should ultimately facilitate the study of tumour tissue bioarchitecture in onconanotomy field.
\end{abstract}


\begin{highlights}
\item AI-driven methods allow refining cell delineation in SEM images
\item A deep cell contour closing operator is proposed 
\item CNN are suitable to fill cell contours in regions with deficient/absent information
\item Low integrity in cell boundaries can be simulated using partial differential equation
\item Cell instance segmentation is enhanced in private and public datasets
\end{highlights}

\begin{keywords}
Segmentation \sep Instance \sep Cell \sep Scanning electron microscopy \sep Deep neural network 
\end{keywords}

\maketitle

\section{Introduction}
\label{Introduction}

In recent years, there has been a growing interest in studying tissue architecture, particularly in the field of oncology, where a deeper understanding of intracellular and intercellular ultrastructure could lead to major advancements in tumour therapy \citep{OmicSpatial, nature_mito, SegNucleiNucleoli, multi-pronged}. \Acf{SEM} has become a crucial imaging technique for providing nanometer-resolution views of the ultrastructure of whole cells and tissues \citep{SBFSEM, Riesterer}.
This technology enables the visualization of organelles such as nuclei, mitochondria, nucleoli, and lipid droplets and so on, but also the study of the organization of cells into the tissue, an essential step for understanding tumour tissue development mechanisms and drug resistance. 
Therefore, the analysis of bioarchitectural parameters within and among tumour cells holds great promise in onconanotomy field, paving the way for a better understanding of tumour tissue 3D organization \citep{1stpaper}.

To this end, an instance segmentation of \ac{SEM} images is mandatory, which involves identifying and separating individual structures, including detecting the boundaries of each structure and assigning a unique label to each.
Each organelle or cellular component possesses distinct characteristics, including number, size, shape, texture, or grey level in images. Consequently, their individual segmentations present diverse methodological challenges \citep{1stpaper}. Manual segmentation is laborious, necessitating automated methods to comprehend the functioning of a heterogeneous lesion in its entirety. Therefore, several studies focused on the development of algorithms for segmenting organelles within cells \citep{SegNucleiNucleoli, MitoEM}. However, cell segmentation and labelling constitute crucial data for collecting cell population statistics. To accomplish this, numerous studies have relied on manually segmenting a subset of cells for inter-cell comparison \citep{OmicSpatial}. However, in particular and unlike organoids, tumour tissue is subject to a great deal of pressure, which means that cells are often tightly packed together. In addition, depending on the tissue fixation method and microscope parameters used, cell membranes are more or less visible. 
Therefore, enhancing computer-aided instance segmentation of closely spaced cells presents a significant challenge addressed within this paper.

Recently, \acf{A.I.} has proven to be an extremely powerful tool for facilitating segmentation \citep{SegHeLaCells, DLSegSurvey}. Among the significant advances, the U-Net architecture introduced by \citet{Ronneberger} has attracted a great deal of interest, particularly in image segmentation. A notable enhancement to U-Net is the use of weight maps in the loss function to more accurately account for cell boundary details, helping to improve segmentation accuracy. 
\citet{DistanceMap_Multiclass} introduced two innovative weight maps integrated into the weighted cross-entropy loss function, which consider the geometric characteristics of cell contours. The shape-aware map assigns substantial weights to narrow and concave areas, leading to superior instance segmentation outcomes. Additionally, by incorporating a third class representing touching borders, they improved the discriminative capability of the network. 
Alternative methods utilize a combination of techniques, integrating deep learning, optical flow, and cross-slice linking, to extrapolate segmentations or manual corrections beyond the original plane \citep{1stpaper, multi-pronged, Seg2Link}. 
More recently, \citet{nnUNet} presented a self-configuring method for biological and medical image segmentation named nnU-Net. This deep learning open-source algorithm automatically configures itself, including preprocessing, data augmentation, network architecture, training and post-processing for any new task and surpasses most existing approaches on many public datasets with various dimensions. Additionally, \citet{nnUNetRevisited} have reaffirmed the continued prominence of \acf{CNN} and nnU-Net as the leading techniques in medical segmentation. 
For the specific case of cell instance segmentation, \citet{3DCellSeg} proposed a method which is conducted using a super voxel-based clustering algorithm. Moreover, they provided an overview of the existing deep learning model for 2D and 3D cell segmentation and lists their major drawbacks. 
Recently, a general-purpose algorithm called Cellpose has been developed and continuously updated to achieve instance cell segmentation across various microscopy modalities and fluorescent markers \citep{CellPose, CellPose2.0, CellPose3}. The authors proposed a vector flow representation that can be predicted by a neural network. Then, by running a gradient ascent procedure for all pixels in an image and checking which pixels converged to the same fixed point, they can assign different label to each cell.

While the aforementioned end-to-end methods show strong efficiency in segmenting cells and assigning a unique label to each, a challenge inherently persists in practice in areas where cell contours are not straightforward, even for a biological expert. In practice, addressing these issues involve filling gaps manually in cell contour segmentation. 
To address the challenge of gap filling, existing approaches in the literature can be broadly classified into three categories: mathematical morphology, variational methods, and deep learning-based techniques. Each of these classes is detailed below: 
\begin{itemize}
\item Mathematical Morphology: In this approach, gap filling in contour objects is typically performed using morphological operations, such as the closing operation, which combines dilation and erosion \citep{Morphological}. However, these methods demonstrate limitations when the gaps are large, elongated, or numerous. In such cases, these methods often alter the overall structure, making them unsuitable for intricate or densely populated regions. Moreover, precise adjustment of the parameters of the morphological filter kernels is generally required, which proves challenging in practice. 
\item Variational Methods: This approach models gap filling as an inpainting problem, frequently utilizing diffusion \acfp{PDE}, such as the heat equation \citep{Inapainting_HeatEq} and the Cahn-Hilliard equation \citep{CH_1stpaper, CH_1stpaper_math, CH_GrayScaleImages}. Additionally, \cite{TV} provides three inpainting codes using heat equation, \ac{TV} inpainting and \ac{TV}-H$^{-1}$ inpainting. 
These models are commonly applied to restore damaged images, such as fingerprint images or grayscale photographs with small inpainting areas. However, these methods require explicit input of missing regions, which poses a significant limitation in applications involving numerous such regions. 
\item Deep Learning-Based Methods: Recently, morphological deep learning frameworks have emerged, replacing traditional convolutions with non-linear morphological filters \citep{DeepMorphoIntro,DeepMorpho}. These frameworks exhibit enhanced capabilities for analyzing complex shapes and patterns in image data, particularly in applications demanding precise structural interpretation. Other deep learning approaches have achieved notable success in inpainting images with irregular and diverse regions \citep{EdgeConnect,InpaintingReview}. Nonetheless, these methods share similar limitations with the morphology- and variational-based approaches. 
Currently, the most effective methodologies leverage \acp{CNN} capable of autonomously detecting missing line segments and accurately reconstructing them with appropriate thickness and curvature, even in images with extensive omissions \citep{CNN, RootGap, RootGap_GAN}.
\end{itemize}

To enhance the aforementioned techniques for cell instance segmentation, our approach focuses on delineating cell boundaries, and our contributions are threefold:

\begin{enumerate} 

\item An \ac{A.I.}-driven cell contour \acf{COp-Net} is proposed: a dedicated \ac{CNN}-based closing network, embedded in an iterative scheme, is introduced to detect and fill specific gaps, as summarized in Figure \ref{general}. We provide a publicly available Python script to reproduce the results from typical test images, which can also be applied to user-provided data.

\item A novel data simulation approach is proposed to simulate local areas with inadequate or missing information using a fully dedicated \ac{PDE} during training, for which we publicly provide the code online for reproducibility purpose.

\item The proposed approach is assessed for cell instance segmentation using both private high-quality \ac{SEM} images from \acf{PDX} tumour tissues and publicly available datasets.
\end{enumerate}

The remainder of the paper is structured as follows: Section 2 presents the complete cell instance segmentation pipeline, detailing each component. Section 3 analyses the performance on private and public testing datasets, showcasing stability in corrections and a significant increase in the proportion of correctly identified cells. Finally, Section 4 discusses the presented results, evaluates the usefulness of each component, and suggests future avenues for exploration.

\section{Material and methods}

\begin{figure}[!ht]
\centering
\includegraphics[width=\textwidth]{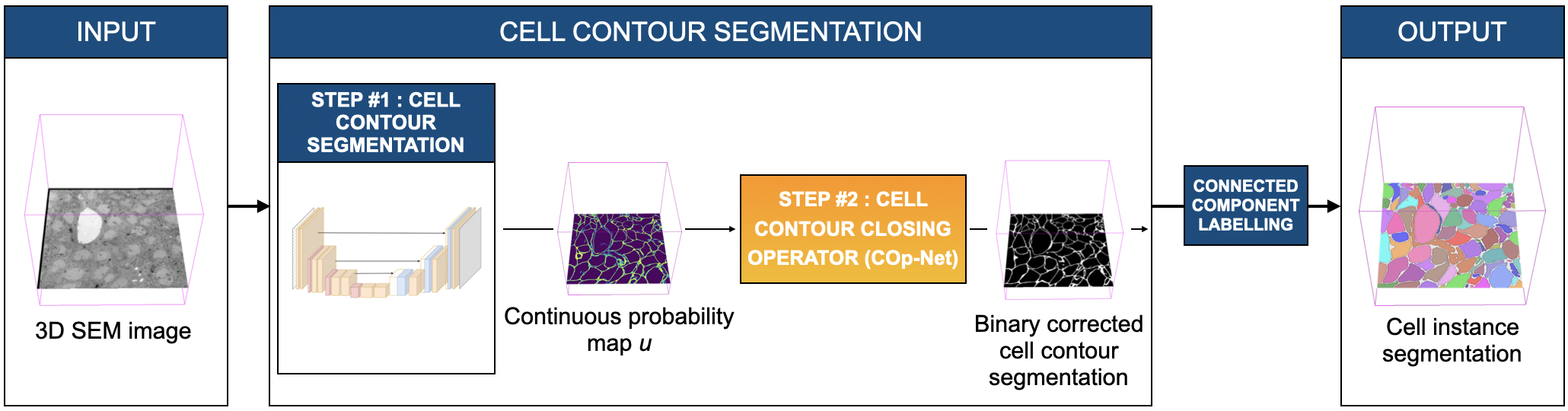}
\caption{Overview of the proposed pipeline for cell instance segmentation. A segmentation neural network first generates a probability map $u$ of cell contours from a 3D SEM stack input (Step \#1, see section \ref{cell contour segmentation}). Subsequently, the proposed \acf{COp-Net} automatically identifies and fills regions with insufficient or missing information (Step \#2, see section \ref{Cell contour closing operator}). Finally, a connected component labelling algorithm is applied to achieve individual cell identification and produce the output cell instance segmentation.\label{general}}
\end{figure}

Our approach for cell instance segmentation relies on the delineation of cell boundaries, which consists of two steps summarized in Figure~\ref{general}. First, a voxelwise cell contour probability map is generated from a 3D \ac{SEM} image (Step~\#1/section~\ref{cell contour segmentation}). A 2D \ac{CNN} \acf{COp-Net} is subsequently applied to address gaps in cell contours on a slice-by-slice basis  (Step~\#2/section~\ref{Cell contour closing operator}). Ultimately, a connected component algorithm is applied to generate the cell instance segmentation.

Note that, our implementation heavily rely on the utilization of the open-source nnU-Net algorithm for both steps due to its self-configuring capabilities, its robust data augmentation, versatile 2D and 3D architectures and performance which align with the standards of leading techniques in medical image segmentation \citep{nnUNetRevisited}.

\subsection{Generation of cell contour probability maps (Step \#1)}
\label{cell contour segmentation}

The nnU-Net algorithm \citep{nnUNet} was trained in a supervised end-to-end manner on ground truth cell contours to produce cell contour probability maps. 
The original implementation of nnU-Net allows saving probabilities by skipping the last activation layer.
In the resulting maps (noted $u$ throughout the rest of the manuscript), each pixel (resp. voxel) value represents the probability of it being part of a cell contour.

\subsection{Deep cell contour Closing Operator (\ac{COp-Net}/Step \#2)}
\label{Cell contour closing operator}

\begin{figure}[!t]
\centering
\includegraphics[width=\textwidth]{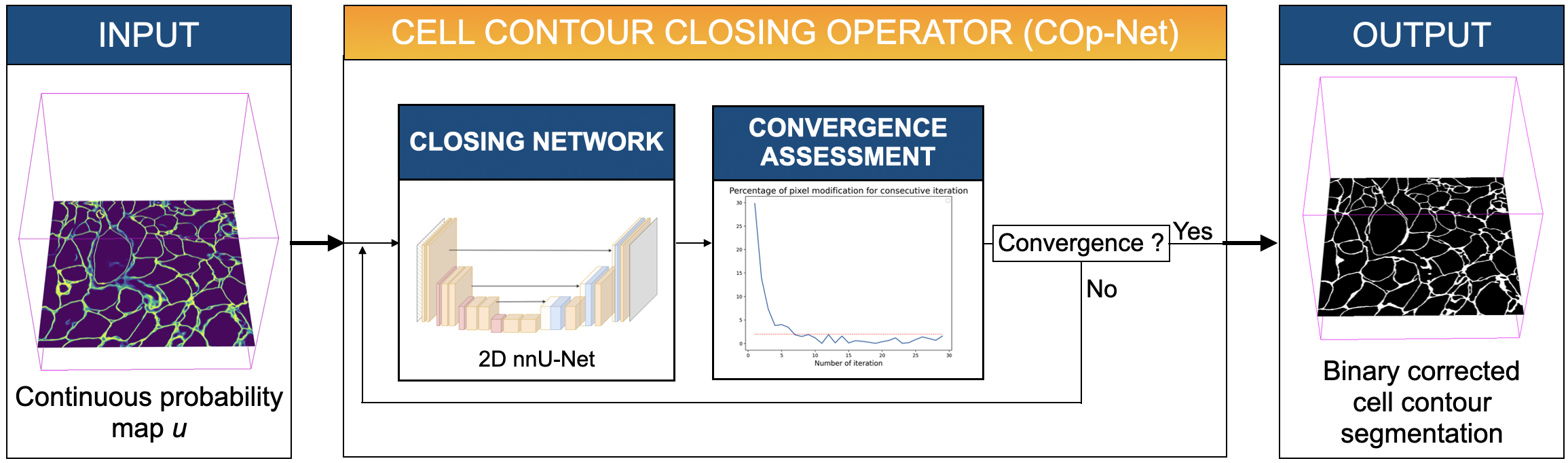}
\caption{Schematic view of the proposed cell contour \acf{COp-Net}. A 2D pixelwise continuous probability map $u$ obtained from Step \#1 is taken as input. The proposed operator is composed by a closing network embedded in an iterative scheme. \label{module}}
\end{figure}

At this point, we have cell contour probability maps $u$ that contain locally inadequate or missing information. The proposed cell contour closing network, \ac{COp-Net}, is then applied to automatically rectify these maps (see Figure~\ref{module}). 
We employed a 2D \ac{CNN} architecture to enable a direct comparison with state-of-the-art cell instance segmentation techniques, which are trained in a 2D context (section~\ref{CellPose} and section~\ref{RootGap}). 
Once trained, the \ac{COp-Net} was applied on a slice-by-slice basis on $u$. 
Once more, we exploited the nnU-Net algorithm \citep{nnUNet}, utilizing its 2D architecture. 
The proposed closing operator is embedded in an iterative scheme with a dedicated convergence criterion. The trained weights for the \ac{COp-Net} cell contour closing network, along with the Python script for iterative \ac{COp-Net} inference, are publicly accessible at \url{https://github.com/Florian-40/CellSegm}.
In the subsequent sections, we elaborate on each component of \ac{COp-Net}.

\subsubsection{Simulating training data using \ac{PDE}}
\label{dilatation}

\begin{figure}[!t]
\centering
  \begin{subfigure}[t]{.33\linewidth}
  \centering 
  \caption*{\ac{SEM} image}
    \includegraphics[width=\linewidth]{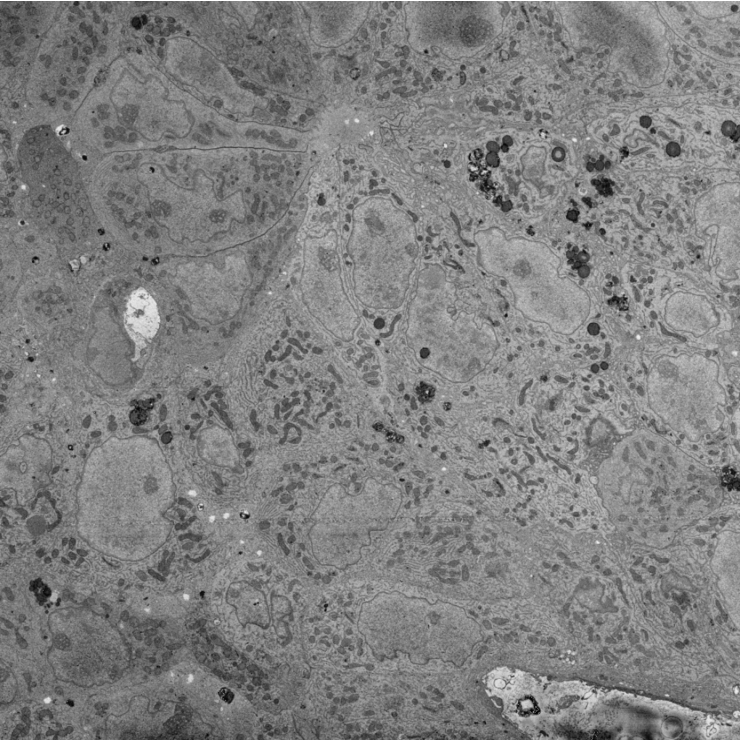} 
    \caption{}
  \end{subfigure}\hfill
  \begin{subfigure}[t]{.33\linewidth}
    \centering
    \caption*{Ground truth cell contour}
    \includegraphics[width=\linewidth]{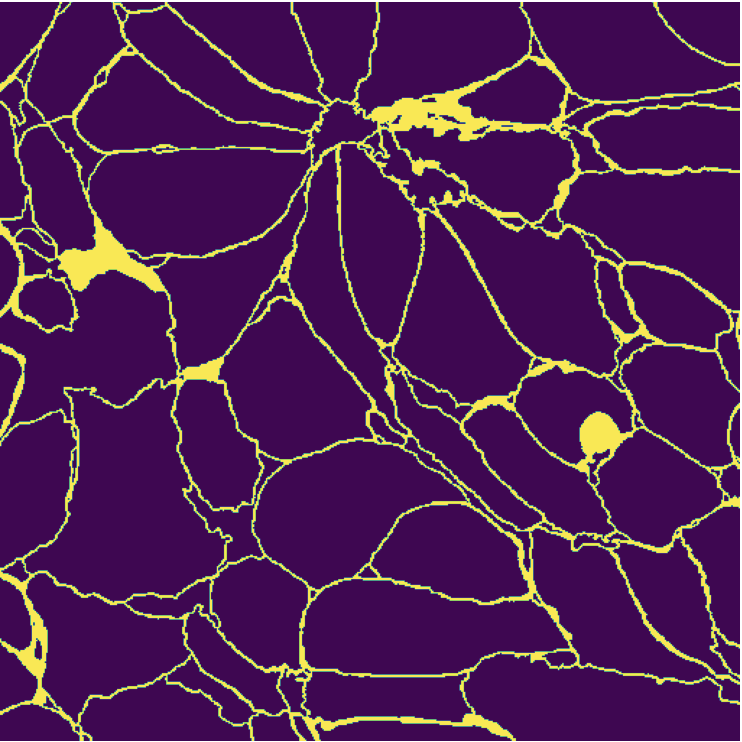} 
    \caption{}
  \end{subfigure}\hfill
  \begin{subfigure}[t]{.33\linewidth}
    \centering
    \caption*{Simulated probability map}
    \includegraphics[width=\linewidth]{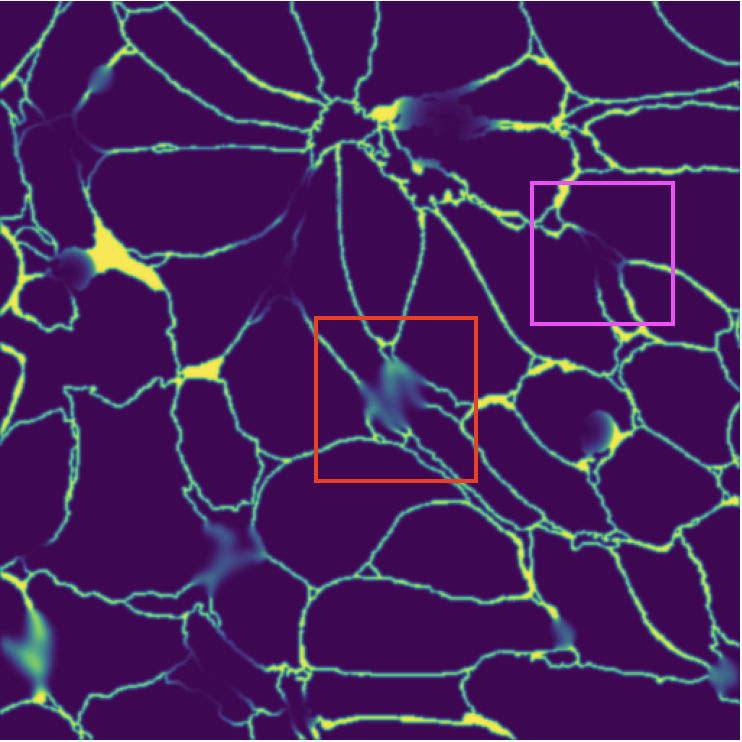} 
    \caption{}
  \end{subfigure}
  
\vspace{0.5cm}  
  \begin{subfigure}[t]{.33\linewidth}
    \centering
    \caption*{Zoom into a local isotropic diffusion area}
    \includegraphics[width=\linewidth]{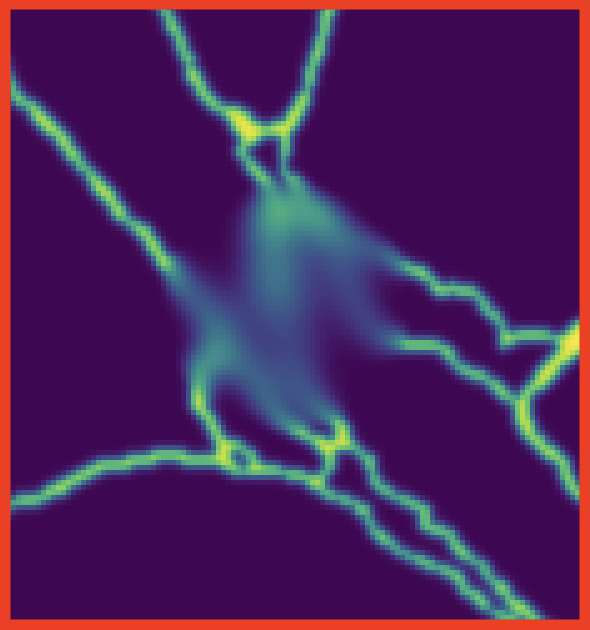} 
    \caption{}
  \end{subfigure}
  \begin{subfigure}[t]{.37\linewidth}
    \centering
    \caption*{Zoom into a local dropout area }
    \includegraphics[width=\linewidth]{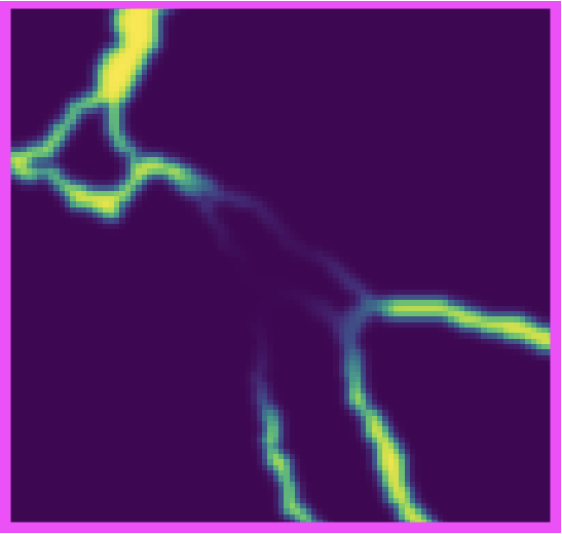} 
    \caption{}
  \end{subfigure}
  \begin{subfigure}[t]{.1\linewidth}
    \centering
    \caption*{}
    \includegraphics[width=.6\linewidth]{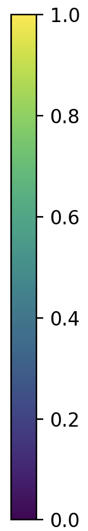} 
  \end{subfigure}
\caption{Typical data used for the end-to-end supervised training of the cell contour closing network. (a): private tumour \ac{SEM} image, (b): corresponding ground truth cell contour , (c): simulated cell contour probability map derived from (b) using Eq.~(\ref{eq1}), (d): zoom into the red square in (c) highlighting a global and local isotropic diffusion of cell contour probability, (e): zoom into the purple square in (c) highlighting a local lack of cell contour probability. \label{slices}}
\end{figure}

The closing network underwent supervised end-to-end training utilizing degraded ground truth cell contour segmentations as inputs to predict their corresponding correct versions. The inputs comprised 2D pixel-wise maps representing cell contour probabilities, incorporating locally partial or missing information generated from ground truth segmentations (see Fig.~\ref{slices}), which were intentionally perturbed using the following \ac{PDE}:


\begin{equation}
\label{eq1}
\begin{array}{cc} 
\partial _t u = \alpha \Delta u - \beta u, \textrm{ in } \Omega & \hspace{2cm}\textrm{(PDE)}\\ 
u(t=0, x) = u_0(x), x \in \Omega & \hspace{2cm}\textrm{(Initial guess)}\\
\vec{\nabla} u \cdot \vec{n} = 0, \textrm{ on }\partial\Omega & \hspace{2cm} \textrm{(Boundary conditions)}\\ 
\end{array} 
\end{equation}

\noindent $\Omega$ being the image domain ($\Omega \subset \mathbb{R}^{2} \rightarrow \mathbb{R}$ for the two dimensional case), $\partial \Omega$ the image boundary, $x \in \Omega$ the 2D pixel coordinates, $n(x)$ is the vector normal to the image boundary at location $x$, $u_0$ the ground truth cell contour segmentation used as an initial guess, $u$ the simulated probability of cell contour at location $x$, and $t$ the scheme time. 
Eq. (\ref{eq1}) was solved using the Crank-Nicolson time scheme and finite differences\footnote{Codes are publicly available at : \url{https://github.com/Florian-40/CellSegm}} \citep{CrankNicolson}. Neumann boundary conditions were applied ($u$ constant along the normal border direction).

While the first contribution in the right part of Eq.~(\ref{eq1}) is tailored to simulate isotropically diffused cell contour probabilities (refer to Fig.~\ref{slices}c and \ref{slices}d), the second contribution addresses local signal drop (refer to Fig.~\ref{slices}c and \ref{slices}e). The first perturbation mechanism is determined by the pixelwise map $\alpha$ ($\alpha \in [d_{min}, d_{max}]$, $d_{min}$ and $d_{max}$ being pre-defined minimal and maximal diffusion values, respectively). The second perturbation is guided by the associated pixelwise perturbation map $\beta$ ($\beta \in [0, 1]$). Both $\alpha$ and $\beta$ maps were randomly generated, and their mathematical formulations are presented in Eq.~(\ref{eq2}):

\begin{equation}
\label{eq2}
\left\{
 \begin{array}{lll} 
 \alpha(x)  &= &d_{min} + (d_{max}-d_{min}) \sum_{i=1}^{N_1}\max \left( 0, 1-\frac{\| \mathbf{x} - \mathbf{x_i} \| _2}{R_i} \right)  \\ 
\beta(x) &=& \sum_{i=N_1+1}^{N_1+N_2}\max \left( 0, 1-\frac{\| \mathbf{x} - \mathbf{x_i} \| _2}{R_i} \right) 
\end{array}
\right.
\end{equation}

\noindent $N_1$ and $N_2$ being the number of local diffusions and the number of local drops in the cell contour probability map, respectively. These local perturbations were applied within distinct circular regions (with radius $R_i$, $\forall i \in \mathbb{N}, 1 \leq i \leq N_1+N_2$) centered on randomly positioned pixels with coordinates $x_i$. The radius $R_i$ were drawn randomly following a uniform distribution $\mathcal{U}\left( \left[\!\left[R_{min},R_{max} \right]\!\right] \right)$ ($R_{min}$ and $R_{max}$ representing  pre-defined minimum and maximum radius values, respectively).

Note that the above method encompass six hyper-parameters $N_1$, $N_2$, $d_{min}$, $d_{max}$, $R_{min}$, and $R_{max}$ for generating random local perturbations to the ground truth cell contour probability map. While some parameters were set empirically to typical values, the main ones were optimized through a grid search analysis, as detailed in section~\ref{hyperparams}.

\subsubsection{Iterative scheme} 
\label{iterative_scheme}

Once trained, the inference of the proposed closing network was performed iteratively, allowing it to progressively detect and fill gaps in cell contours. In each iteration, as the input images evolve, the network refines its predictions by identifying and correcting gaps that may not have been fully captured in a single pass. This iterative approach continues until the percentage of modified pixels between two consecutive iterations fell below a threshold, ensuring the process halts once the correction have converged. A typical threshold of $0.1\%$ was employed to terminate the iterative process.

\subsection{Dataset}
\label{dataset_section}

This section presents a comprehensive overview of the data obtained from a \ac{SBFSEM}, used for training, validating and testing purposes. The proposed approach for cell instance segmentation was assessed using both private high image quality \ac{SEM} \ac{PDX} tumour tissues and publicly available datasets. Additionally, its generalisation to a different imaging context -- nuclei instance segmentation in histopathology images -- is presented in the Supplementary Material.

\subsubsection{Private \ac{PDX} tumour tissue datasets}
\label{sssec:private}

\begin{figure}[!ht]
\centering
\includegraphics[width=\textwidth]{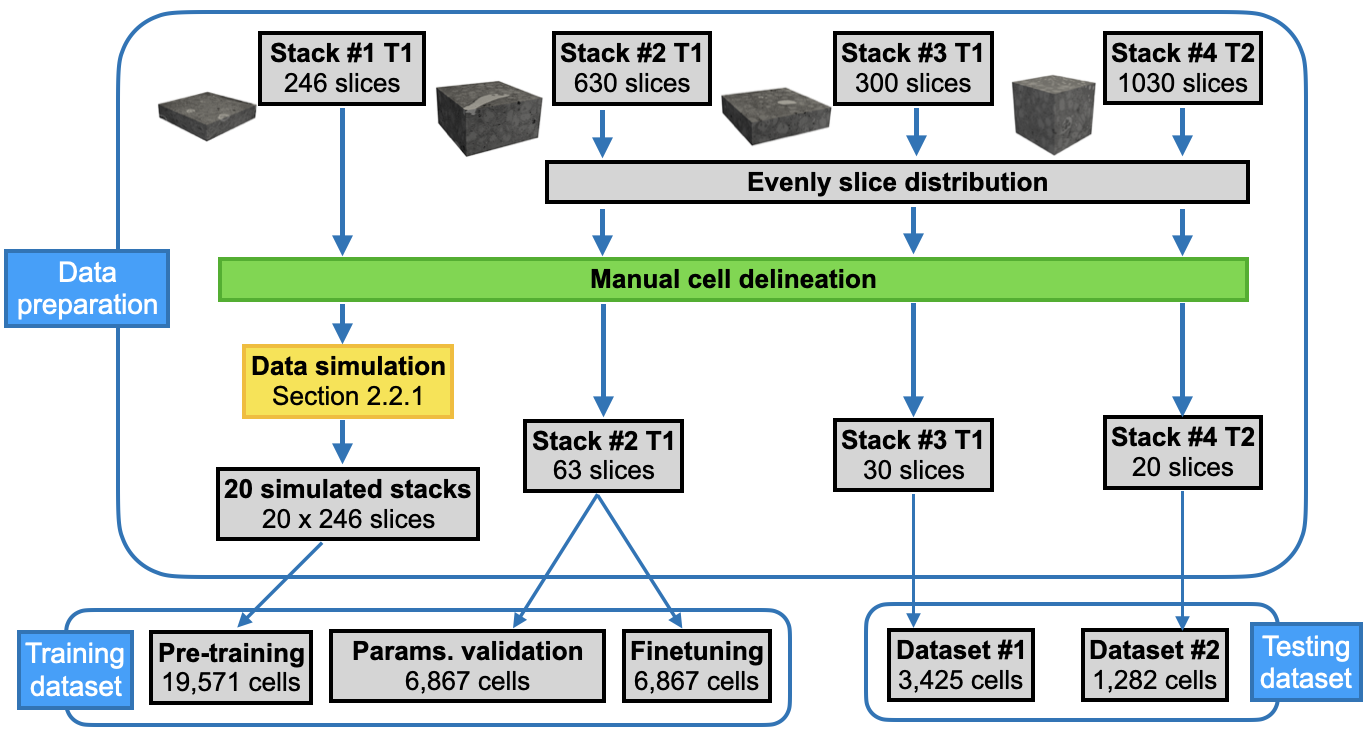}
\caption{Private dataset preparation and distribution used to train, validate and test the \ac{COp-Net}. T1 and T2 represent two different \ac{PDX} tumour tissue samples. The number of cell delineations is specified for both the training and testing datasets. \label{dataset}}
\end{figure}

The private datasets, described in Figure \ref{dataset}, consist of image stacks acquired using \ac{SBFSEM} from two distinct \ac{PDX} hepatoblastoma tissues: one tissue sample (referred to as T1 in the remainder of this manuscript) was designated for training purposes, while the other (referred to as T2) was reserved for testing. For additional information about these \ac{PDX} samples, the reader is referred to \cite{1stpaper}. 

\paragraph{Image pre-processing}
Intensity standardization was accomplished by individually applying z-scoring on a slice-by-slice basis. 
Subsequently, all images were resized to achieve a standardized in-plane dimension of $512 \times 512$ pixels (the value of the new pixel is calculated by considering a weighted average of the pixels that contribute to it in the original image).
Image alignment was completed within each stack to correct for spatial drifts (translations) occurred during \ac{SBFSEM} acquisition \citep{StackReg}. All manual segmentation was done using an interactive drawing tablet and pen (Cintiq pro 5, Wacom) combined with the ITK-SNAP software \citep{ITKSNAP}. For additional information about the image pre-processing procedure, the reader is referred to \cite{1stpaper}.

\paragraph{Training dataset}

The first stack (Stack~\#1 T1, 246 slices, acquisition parameters: voxel size~$= 0.015 \times 0.015 \times 0.1$ \textmu m$^3$, \ac{FOV}~$= 75 \times 75$ \textmu m$^2$) underwent a complete manual segmentation using a semi-automatic method (for additional information about the method employed, the interested reader is referred to \cite{1stpaper}).
These manual segmentations were first employed to train for the initial cell contour segmentation network (Step~$\#1$, section~\ref{cell contour segmentation}). In that case, the 3D architecture of nnU-Net was employed in Step~\#1. Then, the 246 manual 2D segmentations ($19,571$ cell delineations) were employed to generate $20 \times 246$ 2D cell contour probability maps with locally partial or missing information (see section~\ref{dilatation}) to pre-train the closing network (Step~\#2).

\paragraph{Hyper-parameter tuning dataset and fine-tuning dataset}
The second stack (Stack~\#2 T1, 630 slices, acquisition parameters: voxel size~$= 0.015 \times 0.015 \times 0.05$ \textmu m$^3$, \ac{FOV}~$= 90 \times 90$ \textmu m$^2$) belongs to the same tissue as Stack~\#1 (T1) but represents a different region. Stack~\#2 underwent manual segmentation on every tenth slice, resulting in a total of $6,867$ cell delineations across $63$ slices. 
Cell contour probability maps with locally missing information were generated using the previously trained initial cell contour segmentation network (Step~\#1, section~\ref{cell contour segmentation}).
These slices were utilized to validate the hyper-parameter values (see section~\ref{hyperparams}) and to fine-tune the closing network embedded in the proposed \ac{COp-Net} (see section~\ref{Cell contour closing operator}).

\paragraph{Testing datasets}
Two additional stacks acquired from two different tumours were used to test the proposed approach: Stack~\#3 T1 (300 slices) is from the same tissue as Stack~\#1 and Stack~\#2 (T1) but represents a different area, while Stack~\#4 T2 (1,030 slices) is from a distinct \ac{PDX}. 
These stacks share the same acquisition parameters: voxel size~$= 0.015 \times 0.015 \times 0.05$  \textmu m$^3$, \ac{FOV}~$= 90 \times 90$ \textmu m$^2$. Stack~\#3 (T1) underwent manual segmentation on every tenth slice, while Stack~\#4 (T2) underwent manual segmentation on every fiftieth slice. This process yielded $3,425$ and $1,282$ cell delineations in 30 and 21 slices, respectively, providing testing on two different tissues.

\subsubsection{Publicly available HeLa cells dataset}
\label{assess_Hela}

The DIC-C2DH-HeLa cells data\-set \citep{HeLa, HeLa2} was utilized to validate the proposed approach. These 2D images of HeLa cells on a flat glass were acquired to achieve a cell-tracking task over time and are publicly available\footnote{Available at : \url{https://celltrackingchallenge.net/2d-datasets/}}. The Zeiss LSM 510 Meta microscope was used with a pixel size~$=0.19 \times 0.19$ \textmu m$^2$, a \ac{FOV}~$ = 97.28 \times 97.28$ \textmu m$^2$, and a time step of 10 minutes. Each image has a standardized in-plane dimension of $512\times 512$ pixels. 
A first dataset (84 images, $1,112$ cell delineations) was used to train the initial cell contour segmentation network (Step~\#1, section~\ref{cell contour segmentation}). In this  specific context of 2D images, the 2D architecture of nnU-Net was employed in Step~\#1. 
A second dataset (84 images, $1,000$ cell delineations) was used to evaluate our proposed approach. Note that, for a fair comparison with competitive approaches, the proposed \ac{COp-Net} was neither trained nor fine-tuned on either of these public datasets.

\subsection{Experimental setup}

\subsubsection{Assessment of \ac{COp-Net}}
\label{sssec:assess_PDX}

The effectiveness of \ac{COp-Net} for individual cell labelling (\emph{i.e.}, assigning a unique label to each cell) and improving cell contour segmentation (thus reducing the reliance on manual delineation) were both evaluated.

\paragraph{Individual cell labelling} To evaluate the accuracy of cell instance segmentation, we first computed the percentage of cells that were correctly individually labelled. A cell was deemed correctly segmented if the overlap between its mask and the corresponding mask in the ground truth exceeded $85 \%$. Additionally, to provide deeper insights into the necessary corrections, we calculated the percentage of merged cells (false negative contours) and split cells (false positive contours). 

\paragraph{Cell contour segmentation} To evaluate the effectiveness of \ac{COp-Net} in enhancing cell contour segmentation, we calculated the \acf{NSD} and the \ac{clDice}, following the guidelines outlined in the generic framework for metric selection proposed by \cite{SCORES}.

\paragraph{Statistical tests} Statistical analysis of each metric was performed using asymmetric Mann-Whitney U Test \citep{MannWhitney}. Results were deemed statistically significant when the $p$-value fell below the conventional threshold of $0.05$.

\paragraph{Practical assessment of the proposed approach}

To evaluate the practical effectiveness of our approach for biologists, an expert biologist estimated the time required to manually define and correct cell contours using an interactive drawing tablet and pen. Time measurements were performed on a set of randomly selected slices ($n>10$) from the private dataset (testing dataset~\#2) to estimate the typical time required for a single slice to: 1) perform fully manual cell segmentation, 2) correct contours generated by nnU-Net, and 3) correct contours generated by nnU-Net~+~\ac{COp-Net}.

\subsubsection{Hyper-parameter analysis}
\label{hyperparams}

We recall that the approach proposed for simulating training data relies on six hyper-parameters. The two primary ones, $N_1$ and $N_2$, which respectively represent the number of local diffusion and local drop in cell contour probabilities, were evaluated through a grid search across the domain $\{0,\ 6,\ 12\} \times \{0,\ 10,\ 20\}$ in the private \ac{PDX} training data described in section \ref{sssec:private}.  More specifically, the closing network was trained on data simulated from Stack~\#1 using various values of \(N_1\) and \(N_2\), and evaluated on Stack~\#2 using the aforementioned assessment metrics (section \ref{sssec:assess_PDX}). Default values of 6 and 10 were employed for $N_1$ and $N_2$, respectively. The four remaining hyper-parameters in Eq.~(\ref{eq2}) have been set to typical values for local disturbances, as follows: ${R_{min} = 2}$ \textmu m, $R_{max}=7$ \textmu m, $d_{min}=0.1$ \textmu m$^2$.s$^{-1}$ and $d_{max}=1$ \textmu m$^2$.s$^{-1}$.

\subsubsection{Comparison with a competing cell instance segmentation algorithm}
\label{CellPose}

We compared our approach with Cellpose, a state-of-the-art generalist algorithm for cellular segmentation across various image types, which has been continually enhanced and supported to date \citep{CellPose, CellPose2.0, CellPose3}.
This 2D \ac{CNN}-based algorithm produces vertical and horizontal vector flow representations along with a probability map that classifies pixels as either inside or outside cells. Cell labelling is accomplished by applying a gradient ascent procedure to pixels where the probability exceeds 0.5. Cell contour segmentation was also achieved using a specialized function from the \texttt{SimpleITK} library \citep{SimpleITK} applied to the probability map generated by Cellpose.

To ensure a fair comparison with our approach, the latest publicly available ``cyto3'' model was fine-tuned using the segmented slices from Stack~\#1 and Stack~\#2 (which are detailed in section \ref{sssec:private}).

\subsubsection{Comparison with a competing gap inpainting approach}
\label{RootGap}

We also compared our approach with a deep inpainting model designed for detecting and filling gaps in binary root segmentations \citep{RootGap}. Their architecture is built on a 2D fully convolutional encoder-decoder network with skip connections, similar to U-Net. 
Since the network input requires binary cell contours and to ensure a fair comparison with our approach, an adapted data simulation is proposed. Cell contours were indeed removed in $16$ randomly located circles with random radii (\textmu m) following a uniform distribution $\mathcal{U}[1 , 3.5]$ in the manually segmented Stack~\#1, generating $20 \times 246$ pre-training images. Then, the network was finetuned using the segmented slices from Stack~\#2.

\subsection{Hardware and implementation details} 

All training and model evaluations have been completed on a computer running Ubuntu 20.04.6 LTS with the Intel(R) Xeon(R) Gold 6248R CPU and a NVIDIA RTX A6000 GPU with 48 GB of VRAM. 
All implementation were performed using \texttt{Python 3.9}. The \texttt{OpenCV v4.9.0.80} library was employed to resample in-plane images (see section \ref{sssec:private}). The StackReg module \citep{StackReg} within the \texttt{pystackreg v0.2.7} library was used for aligning the stacks on a slice-by-slice basis (see section \ref{sssec:private}). The \texttt{nnU-Net v2.3.1} was employed and publicly available\footnote{Available at : \url{https://github.com/MIC-DKFZ/nnUNet/tree/v2.3.1}}. The \texttt{Cellpose v3.0.8} was employed and publicly available\footnote{Available at : \url{https://github.com/MouseLand/cellpose/tree/v3.0.8}}. \texttt{SciPy v1.12.0} library was employed to compute statistical tests. \texttt{SimpleITK v2.2.1} library \citep{SimpleITK} and \texttt{Scikit-image v0.22.0} were employed to generate connected component labels (see Fig. \ref{general}).

\section{Results}

\subsection{Private \ac{PDX} tumour tissue datasets}

\begin{figure}[!t]
 \begin{subfigure}[b]{.32\linewidth}
    \centering
    \caption*{\ac{SEM} image}
    \includegraphics[width=\linewidth]{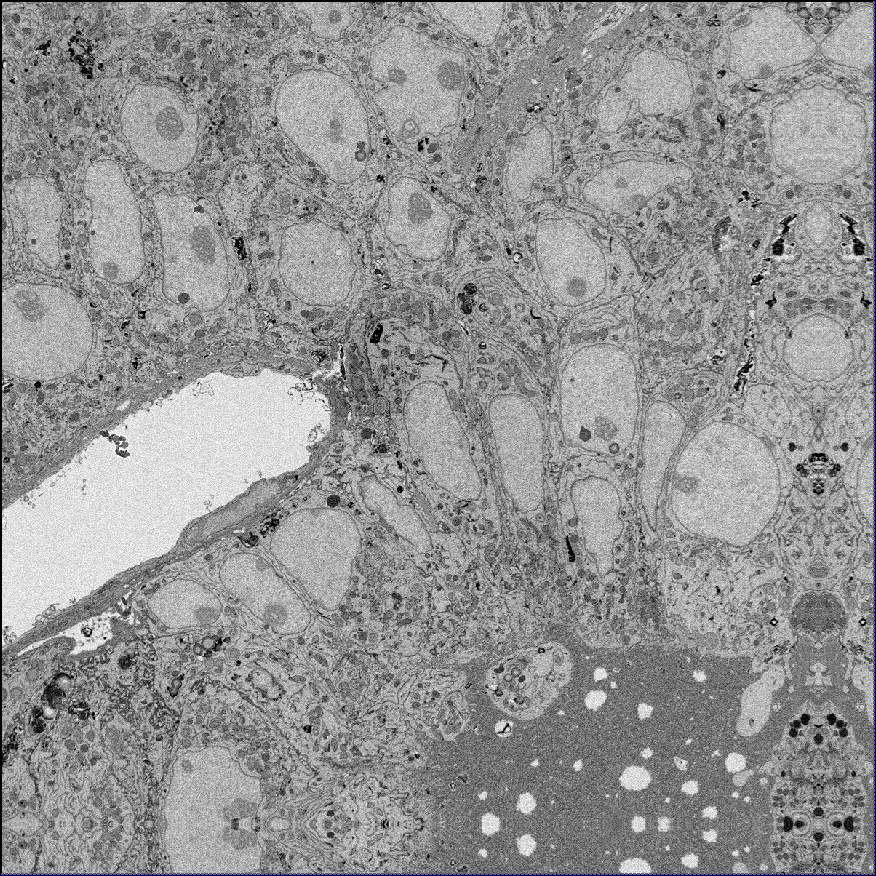} 
    \caption{}
  \end{subfigure} \hfill
  \begin{subfigure}[b]{.32\linewidth}
    \centering 
    \caption*{Ground truth}
    \includegraphics[width=\linewidth]{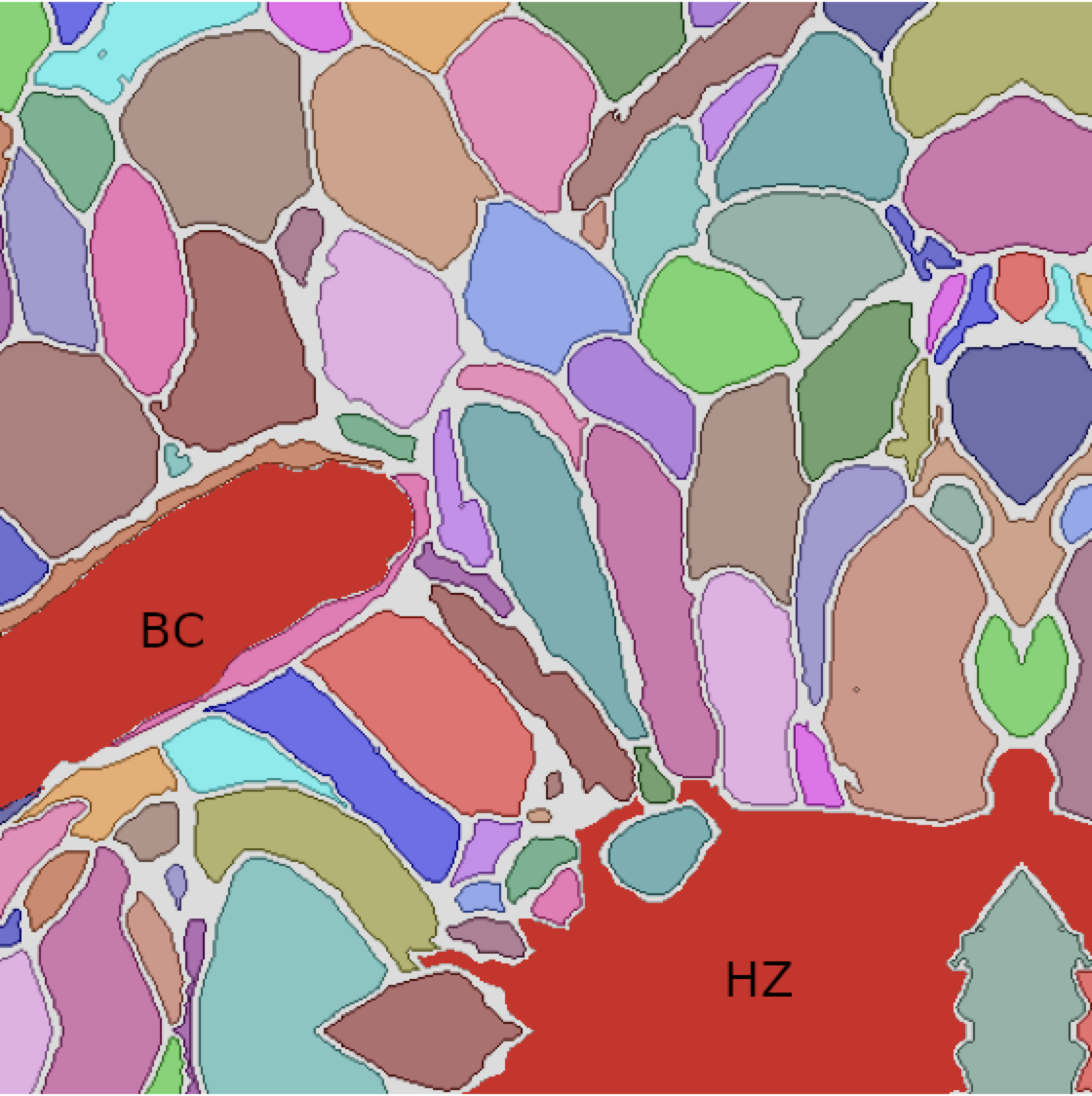} 
    \caption{}
  \end{subfigure} \hfill
\begin{subfigure}[b]{.32\linewidth}
    \centering
    \caption*{Cellpose}
    \includegraphics[width=\linewidth]{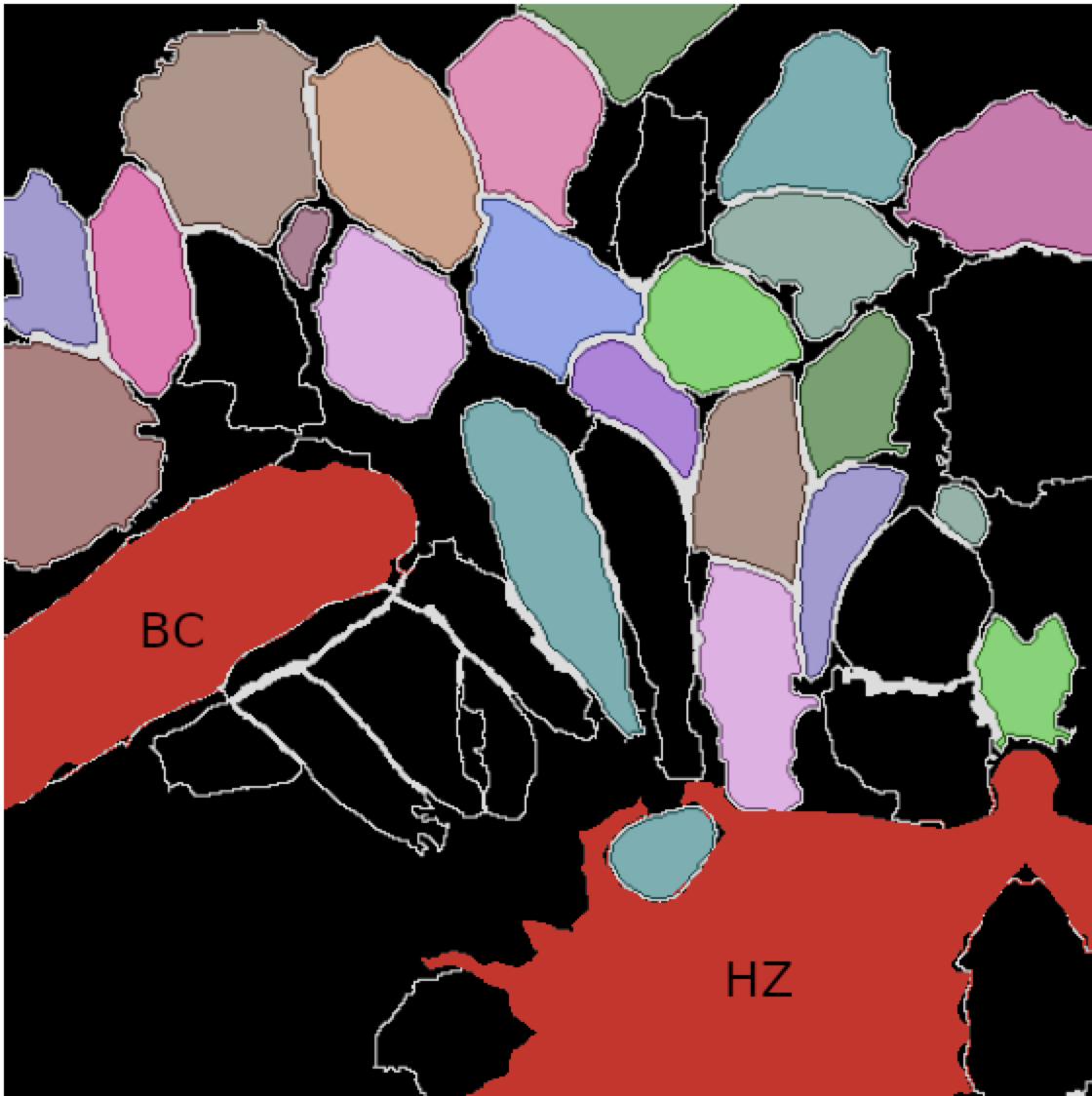} 
    \caption{}
  \end{subfigure}
\begin{subfigure}[t]{.32\linewidth}
    \centering 
    \caption*{nnU-Net}
    \includegraphics[width=\linewidth]{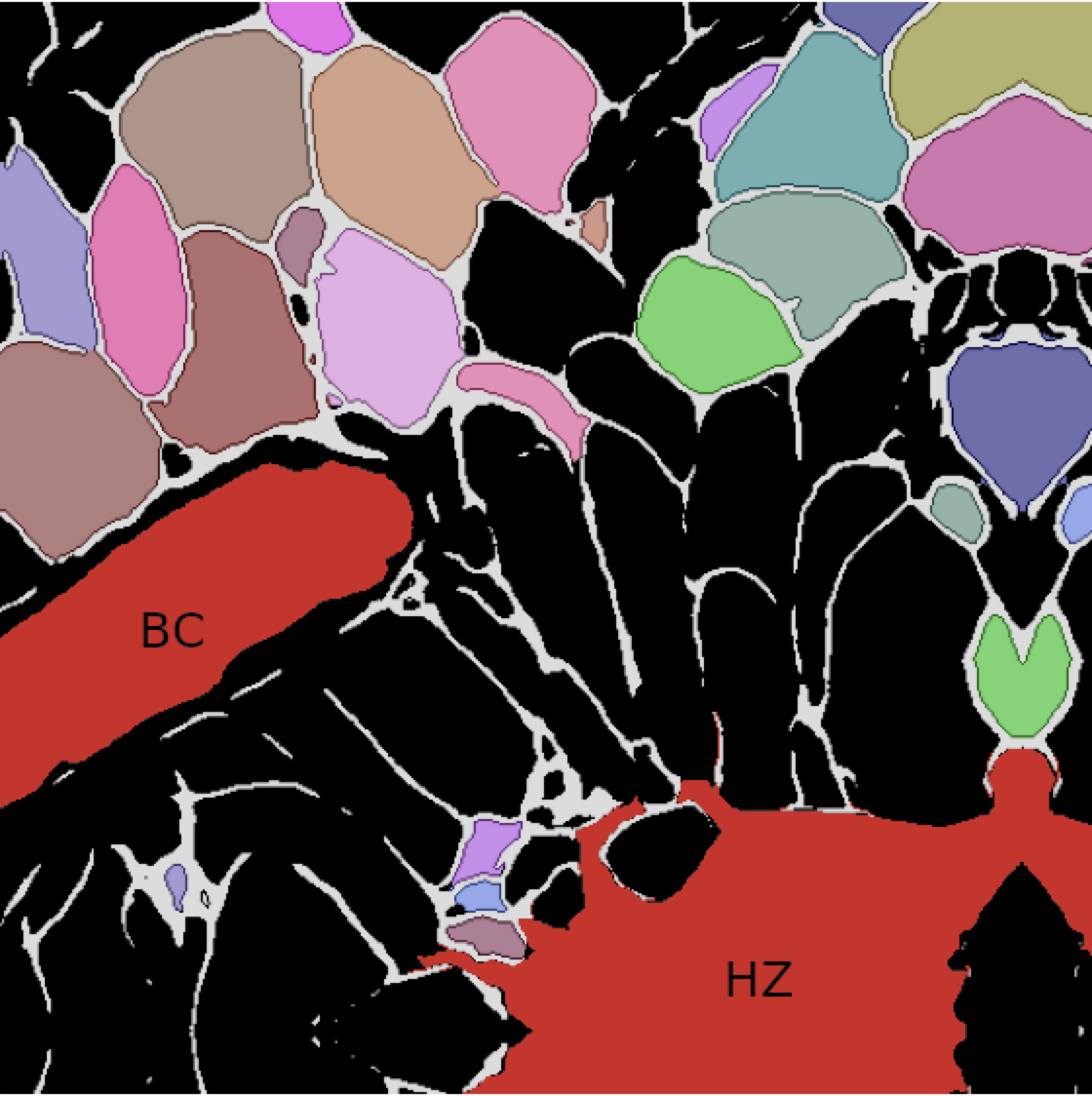} 
    \caption{}
  \end{subfigure} \hfill
  \begin{subfigure}[t]{.32\linewidth}
    \centering 
    \caption*{nnU-Net + \citep{RootGap}}
    \includegraphics[width=\linewidth]{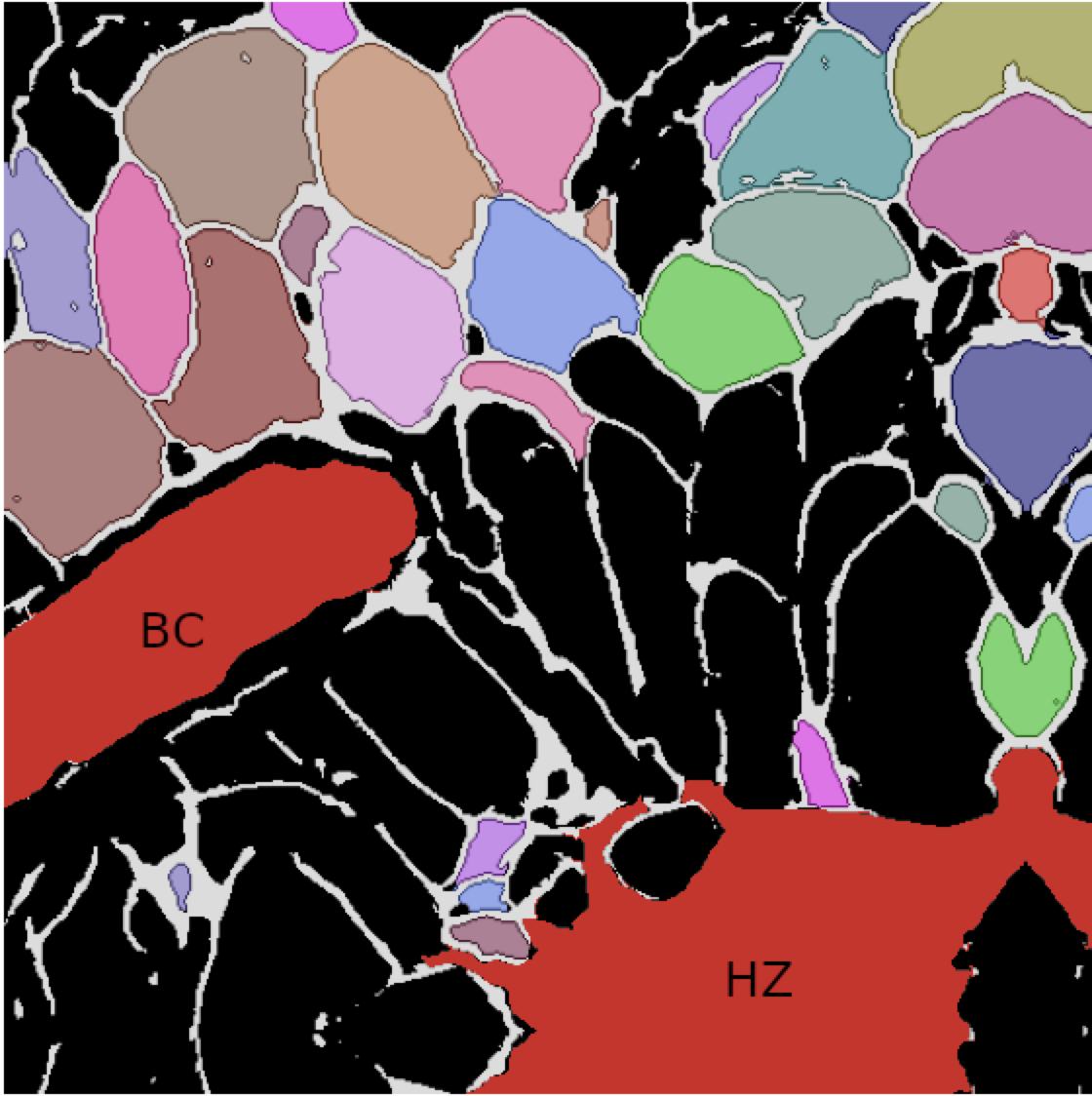} 
    \caption{}
  \end{subfigure}\hfill
  \begin{subfigure}[t]{.32\linewidth}
    \centering 
    \caption*{nnU-Net + \ac{COp-Net}}
    \includegraphics[width=\linewidth]{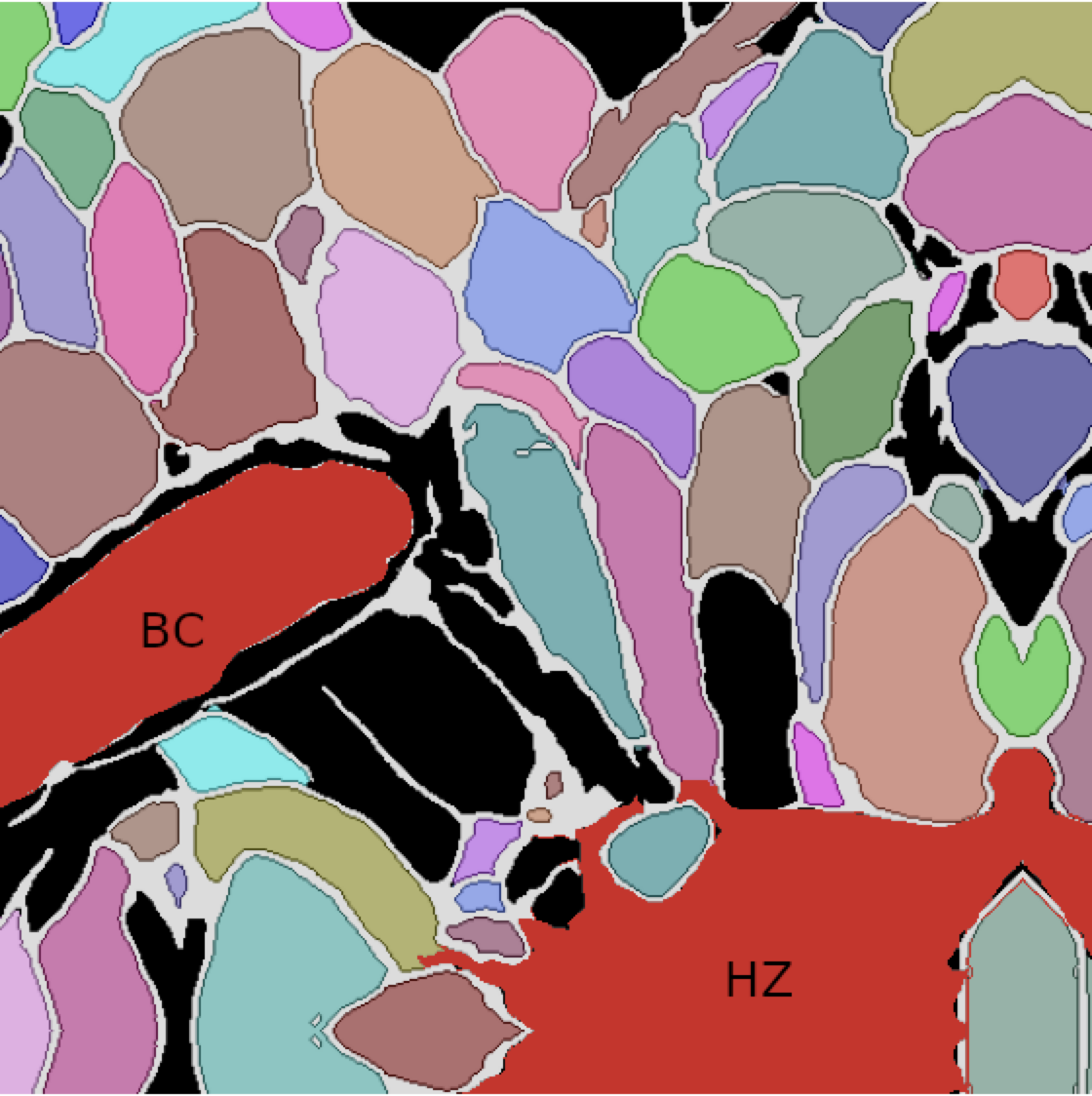} 
    \caption{}
  \end{subfigure}
  \caption{Visual comparison of results obtained on the private testing dataset~\#2. (a) : private \ac{SEM} image, (b-f): white colour represents cell contour, red colour represents blood capillary (BC) and haemorrhagic zone (HZ) obtained through manual segmentation, coloured cells represents accurately labelled cells compared to ground truth (b), and black indicates errors as compared to ground truth, (c): cell contours were obtained using the produced probability map by the Cellpose and a contour function. \label{visu_slices_exter}}
  \end{figure}
  
Figure \ref{visu_slices_exter} provides a visual comparison of the results obtained using the proposed approach and the competing methods on the private testing dataset~\#2 (default hyper-parameters, as defined in section~\ref{hyperparams}). The input \ac{SEM} image (Fig.~\ref{visu_slices_exter}a) and the manual ground truth segmentation (Fig.~\ref{visu_slices_exter}b) are reported, along with results obtained using the Cellpose algorithm \citep{CellPose} (Fig.~\ref{visu_slices_exter}c), the Step~\#1 alone (Fig.~\ref{visu_slices_exter}d), the competing gap inpainting method \citep{RootGap} (Fig.~\ref{visu_slices_exter}e) and the proposed additional module \ac{COp-Net} (Fig.~\ref{visu_slices_exter}f).
In this central slice of the stack, a comparable number of accurately labelled cells can be observed using Cellpose and nnU-Net. While a visual improvement is slightly noticeable with the competing gaps inpainting method \citep{RootGap}, the number of accurately labelled cells further increases using \ac{COp-Net} (Fig.~\ref{visu_slices_exter}f). As expected, segmented cell contours (white pixels) more closely matched the cell boundaries when using approaches involving nnU-Net applied to cell boundaries.

Table \ref{tab:res_private} confirms these visual observations for both testing datasets $\#1$ (3,425 cell delineations) and $\#2$ (1,282 cell delineations). 
The use of \ac{COp-Net} resulted in a substantial improvement in \ac{NSD} and \ac{clDice} compared to the initial cell contour segmentation (Step~\#1), with only a minor increase in the inherent proportion of erroneously split cells (statistically insignificant increase).
It can be noticed that the \ac{clDice} between segmented cell contours in consecutive slices averaged $0.90 \pm 0.05$ (Stack~\#4), indicating consistent correction across successive slices. 
The proposed \ac{COp-Net} significantly enhanced the performances in individual cell labelling up to $72 \%$ and cell contour delineation (\ac{clDice}) up to $0.96$, surpassing the capabilities of the existing state-of-the-art gaps inpainting  and cell instance segmentation approaches.

\begin{table}[!t]
\caption{Assessment of the proposed pipeline (Fig.~\ref{general}) on private tumour tissue datasets ($\#1$=$3,425$ cell delineations, $\#2$=$1,282$). Higher percentages of correctly labelled cells indicate better performance, while lower percentages of erroneously merged or split cells reflect fewer errors. Additionally, higher \ac{NSD} and \ac{clDice} scores correspond to better contour segmentation quality.
Best results are indicated in bold font, and~*~denotes the outcomes of our proposed \ac{COp-Net} that demonstrate significant enhancement over the cell contour segmentation achieved solely by nnU-Net (Step~\#1) ($p < 0.05$).\label{tab:res_private}}
\noindent 
\begin{tabular}{C{3.2cm}C{2cm}C{1.8cm}C{1.5cm}C{1.5cm}C{1.5cm}}
\toprule
&  \multicolumn{3}{c}{\bfseries{Labelling metrics}} & \multicolumn{2}{c}{\bfseries{Segmentation metrics}}\\ 

 & \bfseries Correctly \newline labelled cells \newline [$\%$]&  \bfseries  Erroneously \newline merged cells \newline [$\%$] & \bfseries  Erroneously \newline split cells \newline [$\%$]& \textbf{\ac{NSD}} [A.U.] &  \textbf{\ac{clDice}} [A.U.]\\

\toprule

\multicolumn{5}{l}{\bfseries{Cellpose}} \\

\rowcolor{gray!50}  Testing dataset \#1 &  $42.8 \pm 4.4$ &  $20.5 \pm 4$ &   $36.8\pm 4.4$ &  $0.81 \pm .01$ &  $0.37 \pm .03$ \\
\rowcolor{gray!25}  Testing dataset \#2 &  $29.2 \pm 6.1$&   $ 31 \pm 6.6$&  $39.7 \pm 9.3$&   $0.66 \pm .05$ &   $0.34 \pm .04$ \\

\toprule

\multicolumn{5}{l}{\bfseries{nnU-Net}} \\

\rowcolor{gray!50} Testing dataset \#1 & $61.6 \pm 4.1$ &  $21.7 \pm 4.4$ &  \boldmath  $16.7 \pm 3.3$ & $0.94 \pm .01$ & $0.9 \pm .01$\\
\rowcolor{gray!25} Testing dataset \#2 & $24 \pm 9.1$&  $ 73.5 \pm 9.5$& \boldmath  $2.5 \pm 2.5$&  $0.81 \pm .05$ & $0.80 \pm .04$\\

\toprule

\multicolumn{5}{l}{\bfseries{nnU-Net + \citep{RootGap}}} \\

\rowcolor{gray!50} Testing dataset \#1 & $64.3 \pm 4.1$ & $16.7 \pm 4.2$ & $19.1 \pm 3.4$ & $0.94 \pm .01$ & $0.89 \pm .01$\\
\rowcolor{gray!25} Testing dataset \#2 & $30.7 \pm 8.8$ & $64.2 \pm 9$ &  $5\pm 3$ & $0.82 \pm .04$ & $0.80 \pm .04$\\

\toprule

\multicolumn{5}{l}{\bfseries{nnU-Net + COp-Net (1st iteration)}} \\

\rowcolor{gray!50} Testing dataset \#1 & $70.3  \pm 3.7$ & $11.5 \pm 3.3$ & $18.2 \pm 3.1$ & $0.96 \pm .01$ & $0.91 \pm .01$ \\
\rowcolor{gray!25} Testing dataset \#2 & $59.9 \pm 13.5$ & $31.1 \pm 12.8$ & $8.9 \pm 3.7$ & $0.93 \pm .02$ & $0.90 \pm .03$ \\

\toprule

\multicolumn{5}{l}{\bfseries{nnU-Net + COp-Net (Full convergence)}} \\

\rowcolor{gray!50} Testing dataset \#1 & \boldmath $72^*  \pm 3.6$ & \boldmath $8.3^* \pm 2.9$ & $19.7 \pm 3.4$ &  \boldmath $0.96 \pm .01$ & \boldmath  $0.91 \pm .01$ \\
\rowcolor{gray!25} Testing dataset \#2 &  \boldmath $71.6^*  \pm 9.1$ & \boldmath  $17.9^* \pm 7.8$ & $10.5 \pm 4.5$ & \boldmath $0.93^* \pm .02$ & \boldmath $0.91^* \pm .03$\\

\bottomrule
\end{tabular} 
\end{table}

\begin{figure}[!h]
  \begin{subfigure}[t]{.33\linewidth}
    \centering
    \caption*{Proportion of correctly labelled cell}
    \includegraphics[width=\linewidth]{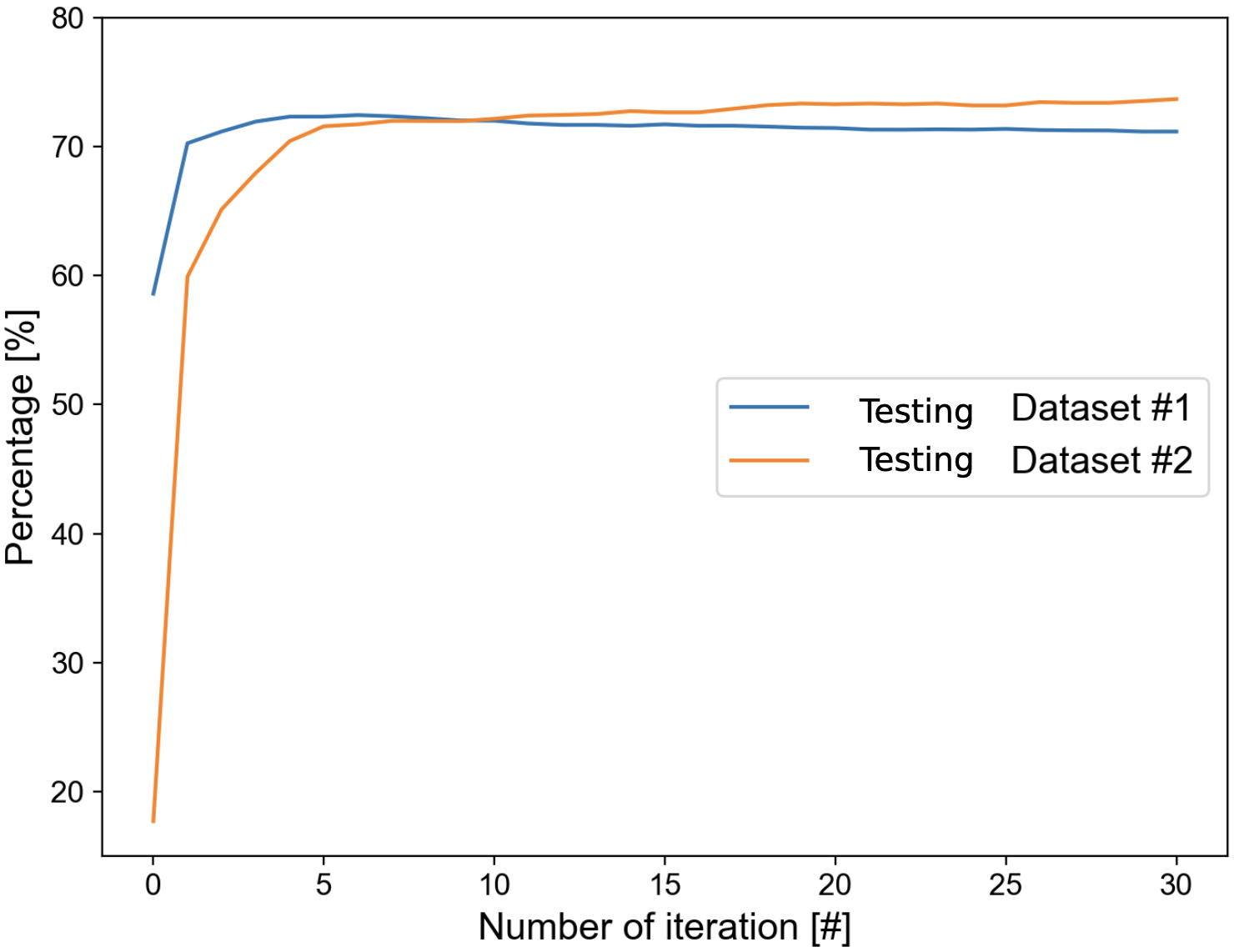} 
    \caption{}
  \end{subfigure}\hfill
  \begin{subfigure}[t]{.33\linewidth}
    \centering
    \caption*{Assessment of cell contour segmentation}
    \includegraphics[width=\linewidth]{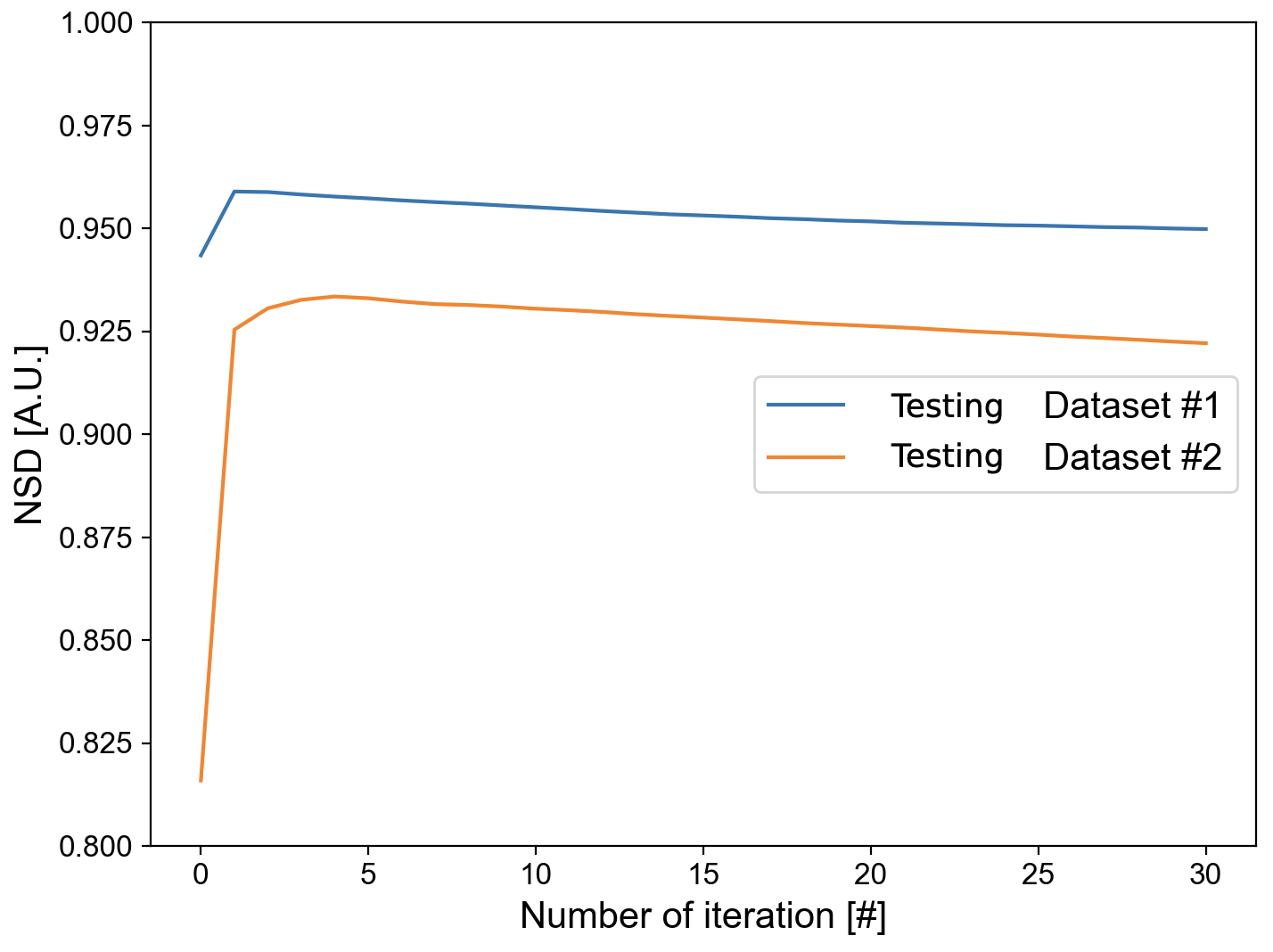} 
    \caption{}
  \end{subfigure}\hfill
    \begin{subfigure}[t]{.33\linewidth}
    \centering
    \caption*{Convergence assessment}
    \includegraphics[width=\linewidth]{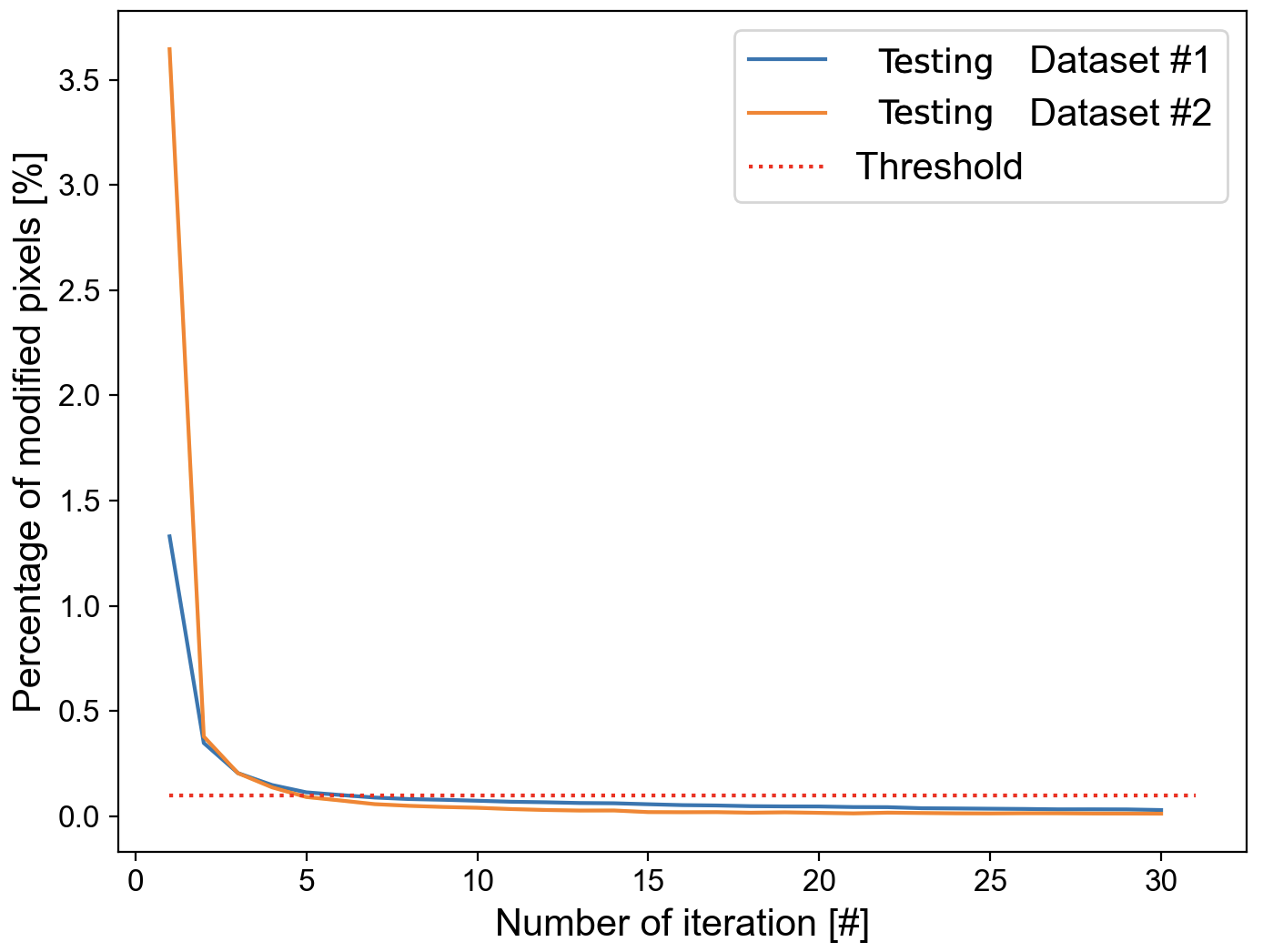} 
        \caption{}
  \end{subfigure} 
  \caption{Result evolution according to the number of iteration of the proposed \ac{COp-Net}. 
  The proposed pipeline was evaluated over 30 iterations on private testing dataset~\#1 (blue) and dataset~\#2 (orange). 
(a): proportion of correctly labelled cells according to the number of iteration, (b): \ac{NSD} to assert the quality of contour corrections, (c): the convergence assessment with a threshold of $0.1\%$ (dotted line) for both testing datasets. \label{graph_res}}
\end{figure}

Figure \ref{graph_res} displays assessment metrics obtained during iterative inference of \ac{COp-Net} on both testing datasets. The proportion of correctly labelled cells primarily improved during the initial iterations (Fig. \ref{graph_res}a) and did not improve significantly thereafter. Using iterative inference of \ac{COp-Net}, the proportion of correctly labelled cells significantly increased from $61.6 \pm 4.1 \%$ (no iteration) to $72 \pm 3.6 \%$ (full convergence/$9^{th}$ iteration) on testing dataset~\#1 (testing dataset~\#2: no iteration~$=24 \pm 9.1\%$, full convergence/$5^{th}$ iteration~$=71.6 \pm 9.1\%$). Conversely, cell contour accuracy deteriorated at some point over iterations, as evidenced by the decline in \ac{NSD} (Fig. \ref{graph_res}b). It can be observed that, in Fig.~\ref{graph_res}c, the convergence threshold of $0.1\%$ for terminating the iterative process effectively balanced the trade-off, optimizing overall performance.


\begin{figure}[!t]
\centering
 \begin{subfigure}[b]{.33\linewidth}
    \centering
    \caption*{\centering Correctly identified cells [$\%$]}
    \includegraphics[width=\linewidth]{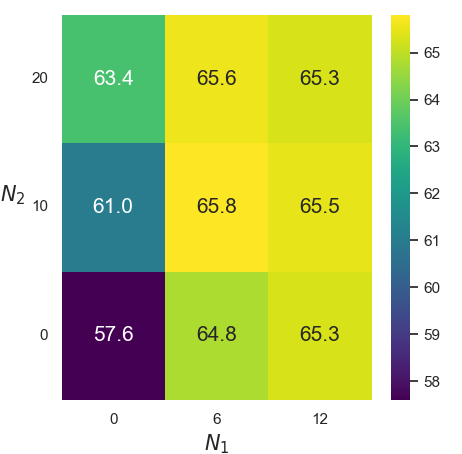} 
    \caption{}
  \end{subfigure}\hfill
  \begin{subfigure}[b]{.33\linewidth}
    \centering
    \caption*{\ac{NSD} [A.U.]}
    \includegraphics[width=\linewidth]{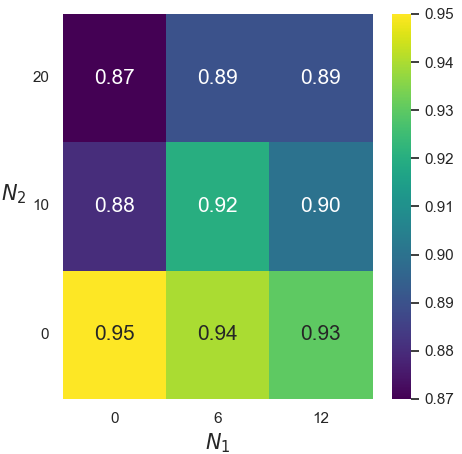} 
    \caption{}
  \end{subfigure}  \hfill
  \begin{subfigure}[b]{.33\linewidth}
    \centering
    \caption*{\ac{clDice} [A.U.]}
    \includegraphics[width=\linewidth]{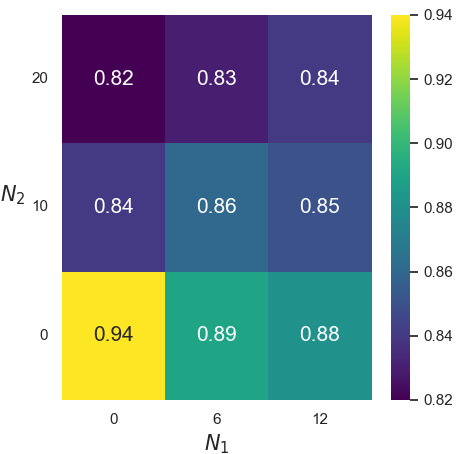} 
    \caption{}
  \end{subfigure}
\caption{Assessment metrics obtained for various values of the hyper-parameters $N_1$ and $N_2$. These hyper-parameters represent respectively the number of local diffusions and local drops in cell contour probability maps. (a): percentage of correctly identified cells, (b): \ac{NSD} (arbitrary unit), (c): \ac{clDice} (arbitrary unit). \label{GridSearch}}
\end{figure}

Figure~\ref{GridSearch} reports assessment metrics obtained during grid-search of hyper-parameters $N_1$ and $N_2$. An insufficient number of perturbation areas in the training data (low $N_1$ or $N_2$ values) resulted in an inadequate detection and filling of gaps, as shown in Fig.~\ref{GridSearch}a. Conversely, highly degraded training data (high $N_1$ or $N_2$ values) resulted in worsened assessment metrics. A balanced trade-off was achieved with $(N_1, N_2) = (6, 10)$ (default hyper-parameters, as defined in section~\ref{hyperparams}).

In our experiments, the estimated average time required for an expert biologist to manually perform cell segmentation was 34 minutes per slice (range: 26 -- 38 minutes) when starting from scratch, 20 minutes per slice (range: 17 -- 24 minutes) when using contours generated by nnU-Net, and 10 minutes per slice (range: 8 -- 15 minutes) when using contours generated by nnU-Net + \ac{COp-Net}. The inference time for \ac{COp-Net} on our test platform was approximately one minute per slice, which represents a negligible computational overhead compared to the significant reduction in manual correction time.

\subsection{Publicly available HeLa cells dataset}

\begin{figure}[!t]
  \begin{subfigure}[b]{.32\linewidth}
  \centering
  \caption*{Grayscale image}
    \includegraphics[width=\linewidth]{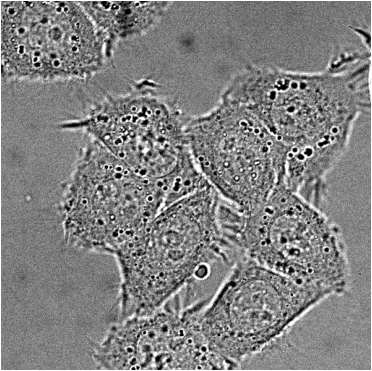} 
    \caption{}
  \end{subfigure}\hfill
  \begin{subfigure}[b]{.32\linewidth}
    \centering
    \caption*{Ground truth}
    \includegraphics[width=\linewidth]{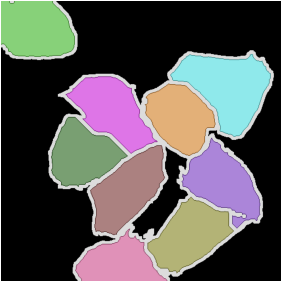} 
    \caption{}
  \end{subfigure}\hfill
  \begin{subfigure}[b]{.32\linewidth}
    \centering
    \caption*{Cellpose}
    \includegraphics[width=\linewidth]{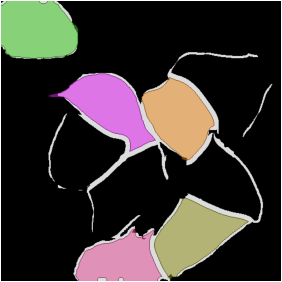} 
    \caption{}
  \end{subfigure}

\vspace{0.5cm}
  \begin{subfigure}[b]{.32\linewidth}
    \centering
    \caption*{nnU-Net}
    \includegraphics[width=\linewidth]{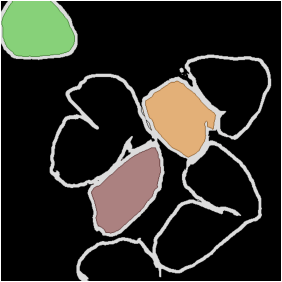} 
    \caption{}
  \end{subfigure}\hfill
  \begin{subfigure}[b]{.32\linewidth}
  \centering
  \caption*{nnU-Net + \citep{RootGap}}
    \includegraphics[width=\linewidth]{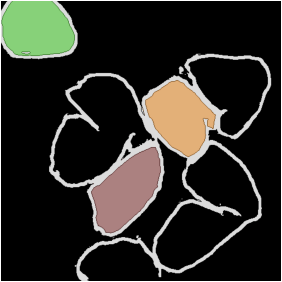} 
    \caption{}
  \end{subfigure}\hfill
  \begin{subfigure}[b]{.32\linewidth}
    \centering
    \caption*{nnU-Net + \ac{COp-Net}}
    \includegraphics[width=\linewidth]{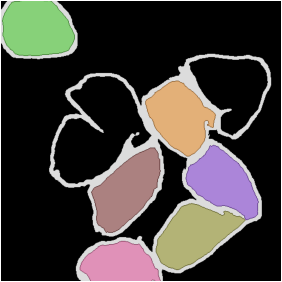} 
    \caption{}
  \end{subfigure}
\caption{Visual comparison of results obtained on the DIC-C2DH-HeLa cell testing dataset. (a): DIC-C2DH-HeLa image \citep{HeLa}, (b-f): white colour represents cell contour, coloured cells represents accurately labelled cells compared to ground truth (b), and black indicates background or errors (except for (b)), (c): cell contours were obtained using the produced probability map by the Cellpose and a contour function\label{visu_slicesHeLa}}
\end{figure}

Figure~\ref{visu_slicesHeLa} provides a visual comparison of the results obtained using the proposed approach and the competing methods on the public testing dataset DIC-C2DH-HeLa (default hyper-parameters, as defined in section~\ref{hyperparams}). The grayscale image (Fig.~\ref{visu_slicesHeLa}a) and the publicly available ground truth segmentation (Fig.~\ref{visu_slicesHeLa}b) are reported, along with those obtained using Cellpose \citep{CellPose} (Fig.~\ref{visu_slicesHeLa}c), the nnU-net (Fig.~\ref{visu_slicesHeLa}d), the competing gap inpainting method \citep{RootGap} (Fig.~\ref{visu_slicesHeLa}e) and the nnU-net supplemented with \ac{COp-Net} (Fig.~\ref{visu_slicesHeLa}f). The competing gap inpainting approach \citep{RootGap} did not improve the initial cell contour segmentation (Fig. \ref{visu_slicesHeLa}e). The highest number of accurately labelled cells and the most accurate segmented cell contours (white pixels) were achieved by supplementing the nnU-Net with the proposed correction module \ac{COp-Net} (Fig. \ref{visu_slices_exter}f).

Table \ref{tab:res_public} confirms these visual observations across the entire testing dataset ($1, 000$ cell delineations). Assessment metrics improved using iterative inference of \ac{COp-Net}: the proportion of correctly labelled cells significantly increased from \(64.2 \pm 17.2 \%\) (no iteration) to \(72.4 \pm 13.8 \%\) (first iteration) and \(74.7 \pm 14.3 \%\) (three iterations), indicating that the most significant improvements occur during initial iterations. It should be noted that the best assessment metrics were achieved with \ac{COp-Net}, together with a minor (statistically insignificant) increase in the proportion of erroneously split cells.

\begin{table}[!t]
\caption{Assessment of the proposed pipeline (Fig.~\ref{general}) on the publicly available dataset ($1,000$ cell delineations). Higher percentages of correctly labelled cells indicate better performance, while lower percentages of erroneously merged or split cells reflect fewer errors. Additionally, higher \ac{NSD} and \ac{clDice} scores correspond to better contour segmentation quality.
Best results are indicated in bold font, and~*~denotes the outcomes of our proposed \ac{COp-Net} that demonstrate significant enhancement over the cell contour segmentation achieved solely by nnU-Net (Step~\#1) ($p < 0.05$).\label{tab:res_public}}
\noindent 
\begin{tabular}{L{4.8cm}C{2cm}C{1.8cm}C{1.4cm}C{1.5cm}C{1.5cm}}
\toprule
&  \multicolumn{3}{c}{\bfseries{Labelling metrics}} & \multicolumn{2}{c}{\bfseries{Segmentation metrics}}\\ 

 & \bfseries Correctly \newline labelled cells \newline [$\%$]&  \bfseries  Erroneously \newline merged cells \newline [$\%$] & \bfseries  Erroneously \newline split cells \newline [$\%$]& \textbf{\ac{NSD}} [A.U.] &  \textbf{\ac{clDice}} [A.U.]\\

\toprule

\bfseries{Cellpose} &  $65.9 \pm 17.8$ &  $30.2 \pm 17.2$ & \boldmath $3.9 \pm 5.5$ & $0.47 \pm .06$ & $0.44 \pm .08$ \\

\bfseries{nnU-Net} & $64.2 \pm 17.2$ & $30.6 \pm 17$ & $5.2 \pm 7.5$ & \boldmath $0.65 \pm .04$ & $0.69 \pm .05$ \\

\bfseries{nnU-Net + \citep{RootGap}} & $68.1 \pm 15.6$ & $25.9 \pm 15.3$ & $6\pm 8$ & $0.65 \pm .04$ & $0.69 \pm .05$\\

\bfseries{nnU-Net + COp-Net (1st iter.)} & $72.4 \pm 13.8$ & $18.1 \pm 14.2$ & $9.6 \pm 9.3$ & $0.64 \pm .04$ & $0.70 \pm .05$ \\

\bfseries{nnU-Net + COp-Net (Full conv.)}& \boldmath $74.7^*  \pm 14.3$ & \boldmath $14.5^* \pm 12.9$ & $10.8 \pm 9.7$ & $0.63 \pm .04$ & \boldmath $0.70 \pm .05$\\

\bottomrule
\end{tabular} 
\end{table}

Finally, the Supplementary Material further substantiates the superior performance of \ac{COp-Net} when generalized to nuclei instance segmentation in histological images.

\section{Discussion}
\label{Discussion}

Our study enhances cell instance segmentation in \ac{SEM} images by integrating a state-of-the-art segmentation technique (nnU-Net) with an innovative \ac{A.I.}-based cell contour closing operator \ac{COp-Net}. The first step of the proposed approach (Step~\#1), based on the nnU-Net framework, enables the calculation of cell contour probability maps containing valuable information that can be utilized by an additional closing operator (Step~\#2). To train the deep closing operator, we propose an innovative simulation strategy based on isotropic diffusion of cell contour probabilities, coupled with local dropout. Obtained results demonstrate the effectiveness of our approach for accurate cellular and nuclei labelling in various imaging context. Moreover, we observed higher \ac{NSD} and \ac{clDice} scores, and a reduced proportion of erroneously merged cells, indicating a significant decrease in false negatives in the output cell contours, thereby reducing the need for residual manual correction.

In our experiments, existing morphological \citep{Morphological, DeepMorphoIntro, DeepMorpho} and variational approaches \citep{Inapainting_HeatEq, CH_1stpaper, CH_1stpaper_math, CH_GrayScaleImages, TV, EdgeConnect, InpaintingReview} for gap inpainting demonstrated suboptimal performance and required extensive hyper-parameter fine-tuning. This limitation rendered them less practical, effectively categorizing them as semi-automatic approaches dependent on user-defined input parameters. On the other hand, to the best of our knowledge, the most recent fully automatic gap inpainting approach was proposed by \cite{RootGap}, which we use as a benchmark for comparison in this study. Similar to \cite{RootGap}, our method addresses the specific challenge of fully automatic cell segmentation. However, our approach offers the following key advancements: 1) the use of probability maps as input instead of binary segmentations, 2) the incorporation of random local perturbations to the ground truth cell contours for training via the PDE in Eq.~(\ref{eq1}), and 3) an iterative strategy that is straightforward to implement and does not require substantial modifications to the nnU-Net codebase.

Two state-of-the-art segmentation approaches serve as baselines for assessing our method. First, it must be underlined that the original nnU-Net represents the current standard for general segmentation tasks in biological and medical imaging and is included in this study. Second, the Cellpose algorithm, which is included in our benchmark, remains the standard for the specific problem of cell instance segmentation. Recent findings \citep{nnUNetRevisited} have shown that newer architectures, such as transformers and Mamba, do not surpass \ac{CNN}-based approaches in segmentation performance, further validating our selection of Cellpose and nnU-Net as comparative methods. That being said, in addition to these fully automatic segmentation approaches, it is worth noting that several methods leverage deep learning to provide efficient semi-automatic segmentation, such as instance segmentation guided by user-provided prompts (e.g., bounding boxes, points, or textual descriptions), including the Segment Anything Model and its derivatives \citep{SAM, SAM2, microSAM}. A direct comparison with these methods is not appropriate due to the fundamentally different nature of the problems they address, which involve user-guided segmentation rather than fully automatic segmentation.

A significant limitation of the Cellpose algorithm lies in its constraints on cell size variability, which contributed to its suboptimal performance.
This hampered the use of this approach with our private \ac{PDX} datasets, which contain 2D slices extracted from 3D datasets, resulting in heterogeneous cell sizes in the obtained in-plane images. Conversely, our approach, which does not impose constraints on cell size variability, proved to be more suitable for our private \ac{PDX} datasets (see Table \ref{tab:res_private}).

When considering the segmentation results obtained with Step $\#1$ across the diverse datasets, it is apparent that generalizing this step is challenging, especially for images acquired from different tissues. The segmentation algorithm relies on grayscale images and learned content, making it sensitive to slight variations even with identical acquisition parameters. Our proposed cell contour \ac{COp-Net} (Step~\#2) exhibits better generalization since it does not depend on grayscale images obtained via \ac{SEM}. Instead, it uses continuous probability maps, leading to more consistent performance on the tested datasets.

While assessment metrics improved with iterative inference of \ac{COp-Net}, cell contour accuracy deteriorated after a certain number of iterations. This trade-off is governed by the convergence threshold used in the iterative inference scheme of \ac{COp-Net} (see section \ref{iterative_scheme}). As illustrated in Figure~\ref{graph_res}, the threshold used to terminate the iterative process was empirically optimized to enhance individual cell differentiation (see Fig.~\ref{graph_res}a) while preserving cell contours (see Fig.~\ref{graph_res}b). 
In our experiments, a threshold of $0.1\%$ yielded $71.6\%$ (private dataset) and $74.7\%$ (public dataset) of correctly labelled cells with \ac{NSD} values of $0.93$ and $0.63$, respectively. These results approach optimal performance of $73.9\%$ (private dataset) and $76.6\%$ (public dataset) of correctly labelled cells, with \ac{NSD} values of $0.93$ and $0.64$, supporting the recommended threshold of $0.1\%$ adopted in this study.

Training data for \ac{COp-Net} were generated by introducing random local perturbations to the ground truth cell contours as described in Eq.~(\ref{eq1}). This methodology enables data augmentation for the training dataset: a total of $246$ manually annotated 2D segmentations were used to produce $20 \times 246$ augmented 2D segmentations for the results presented in this paper. In the proposed nnU-Net-based implementation of \ac{COp-Net}, data augmentation for the training dataset was influenced by two main factors: 1) the magnitude of local perturbations applied to the ground truth cell contours via random gap simulation in Eq.~(\ref{eq1}), and 2) the data augmentation strategies embedded in the nnU-Net framework, which include global transformations such as rotations, scaling, Gaussian noise, blurring, mirroring, and other transformations.
When the extent of local perturbations to ground truth cell contours was increased, the impact of global transformations diminished, and vice versa. Consequently, our experimental results revealed that both increasing and decreasing the number of simulated probability maps to $40$ or $1$ sets led to reduced labelling and segmentation performance. Specifically, the correct labelling rates were $67.2\%$ and $68.9\%$ for $1 \times 246$ and $40 \times 246$ augmented 2D segmentations, respectively (\ac{clDice} = 0.90 for both), compared to $71.6\%$ using $20 \times 246$ augmented 2D segmentations (\ac{clDice} = 0.91) (Table~\ref{tab:res_private}).

Our study has some limitations that will be addressed in future studies:

First, we employed a 2D architecture for the cell contour closing operator \ac{COp-Net} due to the challenges in obtaining manually segmented data and to facilitate comparison with the two aforementioned state-of-the-art techniques. While it is reasonable to expect that performance could be enhanced using a 3D nnU-Net architecture, adopting such an approach introduces additional challenges related to data acquisition, computational memory costs, and processing time. Alternatively, our attempts on the 3D \ac{PDX} datasets for orthogonal slice inferences and 2.5D strategies were not satisfactory due to the interpolations required to correct for significant pixel anisotropy in the out-of-plane direction. Since our proposed module significantly reduces the necessary correction efforts across the datasets at hand, we can now aim to correct entire stacks within reasonable timeframes, thus laying the groundwork for exploring three-dimensional neural networks in future studies. 

Second, the proposed \ac{COp-Net} approach is inherently designed to detect and fill gaps up to a predefined maximum size, set by $R_{max}$ ($R_{max}=7$ \textmu m in the presented results). While large gaps may not be fully resolved in a single inference pass, it is expected that such gaps can still be partially addressed during the process. To mitigate this limitation, the gap-closing network performs inference iteratively: at each iteration, the input to the subsequent inference is a cell contour probability map that evolves from the previous iteration, as cell contours are progressively refined and filled. This iterative strategy is straightforward to implement and does not require extensive modifications to the nnU-Net codebase. While it is theoretically possible to modify the nnU-Net code to update the network itself at each iteration, doing so would necessitate substantial changes to the publicly available nnU-Net codebase.

Finally, it should be emphasized that the performance of the proposed approach stems from the fact that the effectiveness of the closing network (Step~$\#2$) is intricately tied to both the initial segmentation performance and the generation of realistic cell contour probability maps. Further optimization may be investigated in both directions in future work.


\section{Conclusion}

The proposed approach significantly refines cell contour delineation and enhances cell instance segmentation outperforming the capabilities of state-of-the-art cell instance segmentation and gap inpainting approaches. To accomplish this, we integrated a state-of-the-art segmentation technique with an additional deep closing operator to fill cell contours in regions with poor or missing informations. The deep closing operator embeds a dedicated convolutional neural network in an iterative scheme. An innovative training strategy is proposed by generating low integrity probability maps using a tailored partial differential equation, for which we provide the code online.

This study represents a significant advancement in the architectural analysis of tumour tissues, offering a robust tool for cell instance segmentation in 2D and 3D context across multiple sample. Such detailed and accurate segmentation is crucial for the field of onconanotomy, as it allows for precise mapping and understanding of tumour micro-environments. The ability to analyse and compare 3D tissue structures can lead to improved diagnostics and treatment planning, ultimately enhancing patient care in oncology.

\printcredits

\section*{Data availability}

The DIC-C2DH-HeLa cell training and test datasets are publicly available at \url{https://celltrackingchallenge.net/2d-datasets/}. The private \ac{SEM} \ac{PDX} datasets are involved in ongoing studies concerning tumour tissue development mechanisms and drug resistance and, therefore, will not be made publicly available.

\section*{Code availability} 
The source code for simulating cell contour probability maps by solving the PDE in Eq.~(\ref{eq1}), the weights for the closing network, as well as the Python script for reproducing iterative inference in \ac{COp-Net} and assessing metrics, are publicly available at \url{https://github.com/Florian-40/CellSegm}.

\section*{Declaration of Generative AI and AI-assisted technologies in the writing process}

During the preparation of this work, the author(s) used ChatGPT 3.5 in order to rephrase sentences to make them clearer and more precise, as well as to provide improvements for specific wording. After using this tool/service, the authors reviewed and edited the content as needed and take full responsibility for the content of the publication.

\section*{Declaration of competing interest}
Dr C.F. Grosset received a 2-year research grant from Aquitaine Science Transfer and is part of the scientific board of Paediatis.

\section*{Acknowledgement}
Computational analysis was carried out using the PlaFRIM experimental testbed, supported by Inria, CNRS (LABRI and IMB), Université de Bordeaux, Bordeaux INP and Conseil Régional d'Aquitaine (see \url{https://www.plafrim.fr/}). Imaging was performed on the Bordeaux Imaging Center, member of the FranceBioImaging national infrastructure (ANR-10-INBS-04). 

\section*{Funding} 
This work was supported by grants to FR from Inserm and Inria in the frame of the "Santé Numerique" joint program and the Institut National du Cancer (INCa) (N° 2023-018). CFG and MIRCADE also received donations from the following charities: Aidons Marina; Aline en Lutte contre la Leucémie; Cassandra; E.S.CA.P.E.; Eva pour la Vie; Fondation for Addie?s Research; Grandir sans Cancer; Les Amis de Marius; Les Récoltes de l'Espoir; Monaco Liver Disorder, Nathanaël, du Rêve et de l'Espoir; Noëline, ma Fille, ma Bataille; Pour Emma; Scott \& Co; Tous Avec Agosti/Kiwanis International and Warrior Enguerrand. The authors don't have any financial benefit from those Foundations and charities.

\bibliographystyle{cas-model2-names}

\bibliography{samplebibliography-clean}


%
%
%

\end{document}